\documentclass[aps,pre,preprint,longbibliography]{revtex4-1}

\usepackage{hyperref}
\hypersetup{
    colorlinks=true
}

\usepackage[T1]{fontenc}
\usepackage{xr}
\usepackage{amsmath}
\usepackage{multirow}
\usepackage{graphicx}
\usepackage{color}
\usepackage{dcolumn}
\usepackage{bm}
\graphicspath{{IMAGES/}}

\begin{document}

\preprint{APS/123-QED}


\author{Samuel S. Gomez}
\affiliation{Faculty of Mathematics, Natural Sciences, and Materials Engineering: Institute of Physics, University of Augsburg, Universit\"atsstraße 1, 86159 Augsburg, Germany}

\author{Lorenzo Rovigatti}
\email{lorenzo.rovigatti@uniroma1.it}
\affiliation{Department of Physics, {\textit Sapienza} Universit\`a di Roma, Piazzale A. Moro 2, IT-00185 Roma, Italy}
\affiliation{CNR-ISC Uos Sapienza, Piazzale A. Moro 2, IT-00185 Roma, Italy}



\date{\today}

\begin{abstract}

We numerically investigate the dynamics and linear rheology of disordered systems made of patchy particles, focussing on the role of valence, temperature and bonding mechanism. We demonstrate that the dynamics is enslaved to bonding, giving rise to an activated behaviour at low temperature. By independently computing the diffusion constant and the viscosity from the simulations, we also confirm the validity of the Stokes-Einstein relation in valence-limited systems, with two caveats: (i) the diffusion constant requires a finite-size correction, at least at the intermediate density we investigate, and (ii) there is the onset of a breakdown that appears at the lowest temperatures considered. Finally, our results show that the storage and loss moduli of mixtures of divalent and $M$-valent particles exhibit an apparent power-law dependence on frequency, hinting at the possibility of using the composition to finely tune the rheological response of these materials. Our results compare well with literature experimental data on valence-limited DNA nanostars. In addition, the wealth of data we present and analyse here will help to develop and test theoretical frameworks aimed at describing the dynamics of flexible limited-valence particles that self-assemble into disordered networks.
\end{abstract}

\title{Diffusion, viscosity and linear rheology of valence-limited disordered fluids}

\maketitle

\section{Introduction}

Valence-limited systems are composed of particles that can form up to a fixed number of bonds \textit{via} short-range (possibly anisotropic) interactions~\cite{C7CP03149A,Russo_2022}. If this fixed number, which sets the optimal number of neighbours per particle and it is often termed \textit{valence} or \textit{valency}, is smaller than $12$, particles can minimise their bonding energy at densities lower than close packing. As a result, open structures, \textit{i.e.} low-density particle arrangements, are stabilised, leading to the formation of disordered (\textit{e.g.} equilibrium gels~\cite{sciortino2017equilibrium}) or ordered (\textit{e.g.} the diamond or pyrochlore lattices~\cite{PhysRevLett.67.2295,10.1063/1.2206111}) assemblies. As a consequence, limited-valence particles can be used both to model and understand the behaviour of particular atomic and molecular systems (\textit{e.g.} water~\cite{bol1982monte,Russo_2022}), but also to design and realise supramolecular materials with ad-hoc thermodynamic, structural and dynamical properties~\cite{doi:10.1021/la0513611,PhysRevLett.97.168301,smallenburg2013liquids,10.1063/1.4722477,10.1063/1.4849115}.

In the context of valence-limited materials, the most used model system is the patchy particle, originally introduced as a novel class of colloidal particle characterized by distinct regions of attractive and repulsive interactions on the surfaces, resembling patches. In the last two decades, patchy particles have been used to model a wide range of systems, from molecular fluids~\cite{bol1982monte}, to DNA~\cite{doi:10.1021/la063036z}, proteins~\cite{heidenreich2020designer,altan2023patchy}, and even objects interacting only through excluded volume interactions~\cite{harper2019entropic}. In addition, patchy models have also been used to design systems \textit{in-silico} with desired properties. Notable examples are open lattices~\cite{liu2016diamond,PhysRevLett.125.118003,liu2023inverse}, empty liquids~\cite{PhysRevLett.97.168301,10.1063/5.0130808}, non-crystallising fluids~\cite{smallenburg2013liquids,doi:10.1021/nn501138w}, equilibrium~\cite{bomboi2015equilibrium,PhysRevLett.132.078203} and reentrant gels~\cite{10.1063/1.4849115,bomboi2016re}. In most of the disordered cases the basic building block used to encode the limited valence is a DNA nanostar, which is a DNA construct made of $M$ single strands designed to hybridise with each other to form a $M$-valent particle~\cite{biffi2013phase}.

Diffusivity and dynamical arrest of patchy systems have been investigated in the past~\cite{doi:10.1021/jp056380y,10.1063/1.5143221}, and the (micro)rheology of DNA nanostars have been also been measured experimentally~\cite{C8SM00751A,xing2018microrheology,conrad2019increasing}. However, to date there are no comprehensive numerical studies connecting the overall valence with other emerging collective properties such as viscosity and viscoelasticity. Here we fill this gap by evaluating the diffusivity, the viscosity and the Stokes-Einstein relationship that connects them, as well as the linear viscoelastic response in frequency for a wide range of temperatures in patchy systems of varying valence, drawing comparisons with the literature numerical and experimental results. Notably, we enrich the comparison by also considering a swapping mechanism akin to that found in covalent adaptable networks~\cite{kloxin2013covalent}, and that can be readily implemented in DNA-based systems by using toehold-mediated strand displacement mechanisms~\cite{PhysRevLett.114.078104,simmel2019principles}.

\section{Methods}

We simulate pure systems and binary mixtures composed of $N = 1000$ patchy particles. For the pure systems we focus on particles with $M = 2$, $3$ or $4$ patches, while for the mixtures we simulate systems made of $N_2$ divalent particles and $N_k$ particles with $k$ patches, with $k = 3$ or $k = 4$. For the latter we define the fraction of divalent particles, $x_2 \equiv N_2 / N = N_2 / (N_2 + N_k)$, and also the fraction of divalent patches, defined as

\begin{equation}
\label{eq:f2}
f_2 \equiv \frac{2N_2}{2N_2 + k N_k} = \frac{2x_2}{2 x_2 + k(1 - x_2)}.
\end{equation}

Table~\ref{tbl:mixtures} reports a summary of the compositions investigated.

\begin{table}[htp]
\begin{center}
\begin{tabular}{ |c|c|c|c|c|c| } 
 \hline
Type & $\langle M \rangle$ & $x_2$ & $f_2$ & $X_{\rm min}$ for $\lambda = 10$ & $X_{\rm min}$ for $\lambda = 1$\\
\hline
\hline
Pure $M = 2$  &  2  &  1 & 1 & 0.026 & 0.001\\
\hline
Pure $M = 3$  &  3  &  0 & 0 & 0.017 & 0.001\\
\hline
Pure $M = 4$  &  4  &  0 & 0 & 0.011 & 0.001\\
\hline
\hline
Mixture of $M = 2, 3$  &  2.2  &  0.8 & 0.73 & 0.022 & 0.001\\
\hline
Mixture of $M = 2, 3$  &  2.4  &  0.6 & 0.5 & 0.021 & 0.001\\
\hline
Mixture of $M = 2, 3$  &  2.6  &  0.4 & 0.31 & 0.020 & 0.001\\
\hline
\hline
Mixture of $M = 2, 4$  &  2.2  &  0.9 & 0.82 & 0.023 & 0.001\\
\hline
Mixture of $M = 2, 4$  &  2.4  &  0.8 & 0.67 & 0.021 & 0.001\\
\hline
Mixture of $M = 2, 4$  &  2.6  &  0.7 & 0.54 & 0.019 & 0.001\\
\hline
\end{tabular}
\caption{Systems investigated in this work. For each system we specify the average valence $\langle M \rangle$, the fraction of divalent particles $x_2$ and the fraction of divalent patches $f_2$ and the fraction of unbonded patches at the lowest temperature considered, $X_{\rm min}$, for the non-swapping and swapping systems. \label{tbl:mixtures}
}
\end{center}
\end{table}%

Particles are modelled as spheres decorated with attractive patches. The excluded volume acting between two particles whose centres are separated by a distance $r$ is modelled with the Weeks-Chandler-Andersen potential, \textit{viz.}

$$
V_{WCA}(r) = 
    \begin{cases}
      4\varepsilon_{\rm rep} \left[ \left( \frac{\sigma}{r} \right)^{12} -  \left( \frac{\sigma}{r} \right)^{6} \right] + \varepsilon_{\rm rep}, &\, r\leq 2^{1/6}\sigma \\
      0, & \, r> 2^{1/6}\sigma
    \end{cases}
$$

\noindent
where $\sigma$ is the particle diameter and $\varepsilon_{\rm rep}$ controls the strength of the repulsion.

The interaction potential acting between two patches separated by a distance $r_{\rm pp}$ is given by

\begin{equation}
  V_{\rm pp}(r_{\rm pp}) = A\varepsilon \left[ B \left( \frac{\sigma_s}{r_{\rm pp}} \right)^4 -1 \right] e^{\sigma_s/(r_{\rm pp}-r_c)}  
  \label{eq:Vpp}  
\end{equation}

\noindent where $\sigma_s = 0.4 \sigma$ defines the position of the minimum of the potential, $r_c = r_s \sigma_s$ is the distance at which the potential vanishes, $r_s = 1.5$, $B = \frac{1}{1+4\left( 1-r_s\right)^2}$ and $A = -\frac{1}{B-1}\frac{1}{e^{1/(1-r_s)}}$.

\begin{figure}[ht!]
   \centering
   \includegraphics[width=0.45\textwidth]{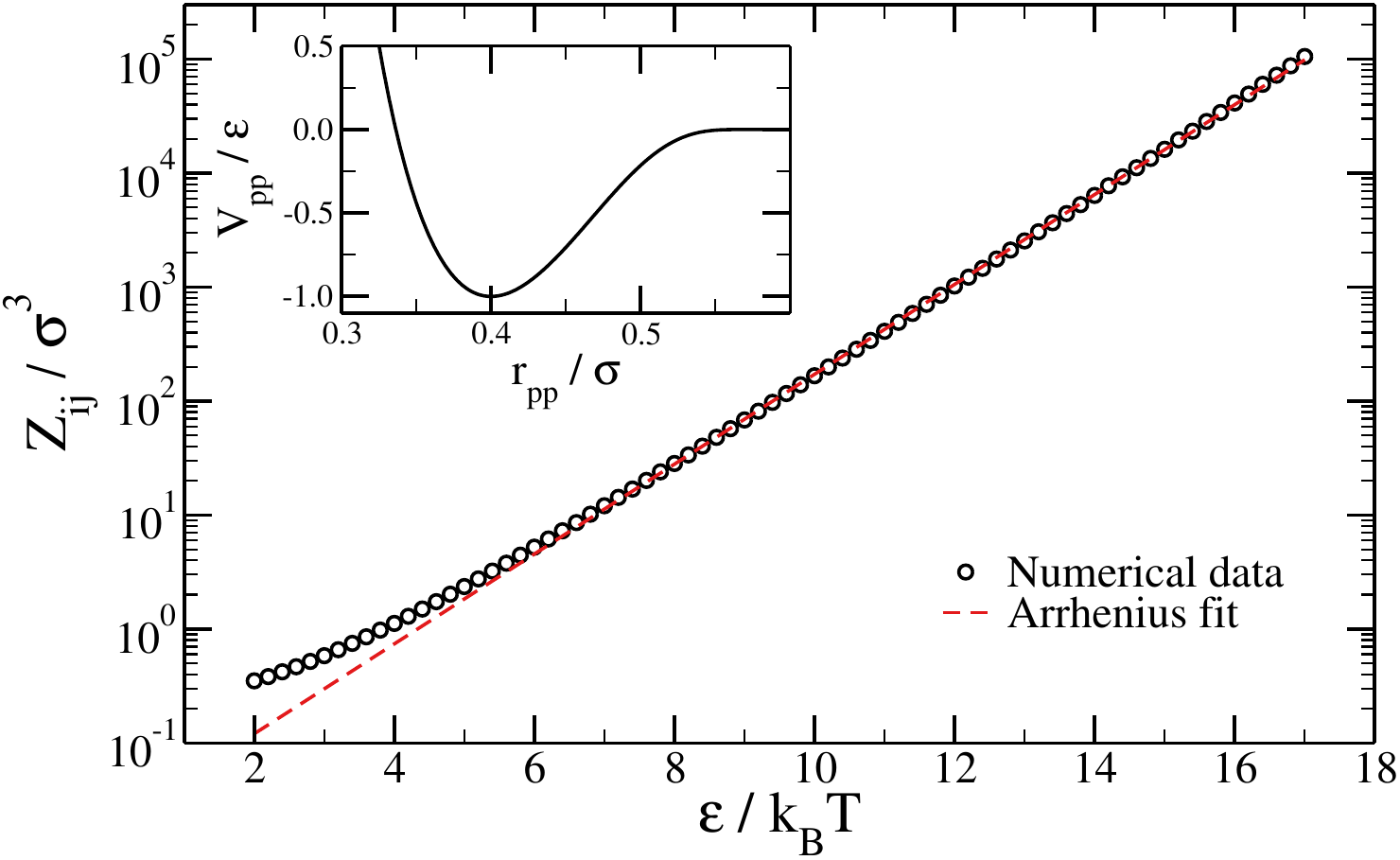} 
   \caption{The dependence on inverse temperature of the dimer partition function. Symbols are data, the line is an exponential fit in the $\varepsilon / k_B T > 6.5$ regime which yields an activation energy $\approx 0.91 \varepsilon$. Inset: the patch-patch interaction potential, Eq.~\eqref{eq:Vpp}.
}
\label{fig:dimer_Z}
\end{figure}

As shown in the inset of Figure~\ref{fig:dimer_Z}, the patch-patch potential resembles a soft square well with a depth of $\varepsilon$. Given a very diluted system of volume $V$, the probability that two one-patch particles, $i$ and $j$, are bonded through this potential is proportional to the dimer configurational partition function, \textit{viz.}

\begin{equation}
Z_{ij} = 4 \pi \int_V e^{-V_{\rm pp}(R_{ij}, \Omega_i, \Omega_j) / k_B T} R_{ij}^2 dR_{ij} d\Omega_i d\Omega_j 
\end{equation}

\noindent where $R_{ij}$ is the distance between the two particles and $\Omega_k$ is the orientation of particle $k$. Computing this quantity numerically yields the results reported in Figure~\ref{fig:dimer_Z}, which shows that at sufficiently low temperature ($\varepsilon / k_B T > 6.5$) $Z_{ij}$ takes on an Arrhenius behaviour with an activation energy $E_Z \approx 0.91 \varepsilon$. As a result, in this regime the patch-patch potential we use can be mapped onto a square well of depth $\varepsilon_{\rm eff} = 0.91 \varepsilon$.

The two-body patch-patch attraction is complemented by a three-body term that enforces the single-bond-per-patch condition even for large patches, thus making it possible to include a bond-swapping mechanism through which patches can exchange bonding partners without breaking any bond~\cite{sciortino2017three}. This potential takes the form

\begin{equation}
V_{\rm 3b} = \lambda  \sum_{i j k} \varepsilon V_3(r_{\rm pp}^{ij}) V_3(r_{\rm pp}^{ik}),
\label{eq:3b}
\end{equation}

\noindent
where the sum runs on triplets of bonded patches $ijk$. The pair potential $V_3(r_{\rm pp})$ is defined in terms of $V_{\rm pp}(r_{\rm pp})$ as
    \begin{equation*}
    V_3(r_{\rm pp})=
    \begin{cases}
      1, &\qquad r_{\rm pp}\leq \sigma_s \\
      -\frac{V_{\rm pp}(r_{\rm pp})}{\varepsilon}, &\qquad \sigma_s\leq r_{\rm pp} \leq r_c.
    \end{cases}
    \label{eq: V3}
    \end{equation*}
    
The $\lambda$ constant in Eq.~\eqref{eq:3b} is a parameter that controls the extent of the bond swapping: for $\lambda = 1$ the energy of a configuration $i-j-k$ where patch $j$ interacts with both patches $i$ and $k$ is approximately the same as a configuration where only the $i-j$ or $i-k$ bonds exist. As a result, either $i$ or $k$ can break free without incurring in any energetic penalty, so that an $i-j$ bond can turn into a $j-k$ bond at constant energy. By contrast, if $\lambda > 1$ the formation of an $i-j-k$ triplet has an energetic cost, and therefore the $i-j$ bond has to break in order for the $j-k$ bond to form. We simulate each combination of system composition and state point with ($\lambda = 1$) and without ($\lambda = 10$) bond swapping.

In the following we will focus on the diffusion constant and the shear viscosity. The former is evaluated as

$$
D = \lim_{t \to \infty} \frac{\langle\Delta r^2\rangle}{6 t}
$$

\noindent
where $\langle\Delta r^2\rangle$ is the single-particle mean-squared displacement averaged over all the particles.

The shear viscosity is estimated in equilibrium simulations as $\eta = \int_0^\infty G(t) dt$, 
where $G(t)$ is the stress autocorrelation function (or shear stress relaxation modulus). We leverage the isotropy of the systems we study and use the following expression for $G(t)$, which is derived by averaging over different directions:

\begin{equation}
\begin{split}
G(t) & = \frac{V}{5 k_B T} \left( \langle \sigma_{xy}(t) \sigma_{xy}(0) \rangle + \langle \sigma_{yz}(t) \sigma_{yz}(0) \rangle + \langle \sigma_{xz}(t) \sigma_{xz}(0) \rangle \right) \\
& + \frac{V}{30 k_B T} \left( \langle N_{xy}(t) N_{xy}(0) \rangle + \langle N_{yz}(t) N_{yz}(0) \rangle + \langle N_{xz}(t) N_{xz}(0) \rangle \right)\\
\end{split}
\label{eq:Gt}
\end{equation}

\noindent
where $\langle \sigma_{ij}(t) \sigma_{ij}(0) \rangle$ is the autocorrelation of the $ij$-th component of the stress tensor at time $t$, $\sigma_{ij}$, and $N_{ij} \equiv \sigma_{ii} - \sigma_{jj}$, with $\langle N_{xy}(t) N_{xy}(0) \rangle$ being its autocorrelation.
We evaluate Eq.~\eqref{eq:Gt} in simulations by using the multi-tau method~\cite{10.1063/1.3491098,10.1063/5.0090540}, and then Fourier-transform the resulting piecewise linear function $G(t)$ to obtain the loss and storage moduli, $G''(\omega)$ and $G'(\omega)$~\cite{10.1063/5.0090540}.

In the following, $p_b$ and $X = 1 - p_b$ are the probabilities that a patch is bonded and unbonded, respectively. Table~\ref{tbl:mixtures} shows that we cool the systems down to temperatures at which less than $3\%$ (for $\lambda = 10$) and $0.1\%$ (for $\lambda = 1$) of the bonds are broken for all the studied compositions. $\varepsilon = \varepsilon_{\rm rep}$ is the depth of the patch-patch attraction and is taken as the unit of energy, $\sigma$ is the diameter of the particles and is taken as the unit of length and time is measured in units of $t_0 = \sigma \sqrt{m / \varepsilon}$, where $m$ is the mass of a particle. For reference, if we consider $\varepsilon \approx k_B T$, we find that $t_0$ is of the order of nanoseconds for DNA nanostars ($\sigma \approx 20$ nm and $m \approx 60k$ Da) and of microseconds for colloids ($\sigma \sim 1$ $\mu m$, $m \sim 10^{-19}$ Kg).

We run molecular dynamics simulations in the microcanical ensemble (\textit{i.e.} at fixed number of particles $N$, volume $V$ and energy $E$). The equations of motion are integrated with a time step $\delta t = 0.001\, t_0$ or $0.002 \, t_0$, depending on the state point. We first equilibrate each system for up to $2 \times 10^8$ time steps by coupling the simulation to a thermostat at a fixed temperature $k_B T / \varepsilon$, and then switch off the thermostat and perform constant-energy production runs for $2 \times 10^9$ time steps. We take averages over $4-8$ independent simulations, depending on the state point, to improve the statistics.

We fix the density $\rho \sigma^3 = N \sigma^3 / V$ at the so-called optimal network-forming density, namely $\rho \sigma^3 = 0.60$, where tetravalent patchy particles can be cooled down without encountering any phase separation~\cite{smallenburg2013liquids} to make it possible to draw a comparison between all the investigated compositions. We note that at this value of the density the effect of hydrodynamic interactions, which are not included in our description, should be small.

\section{Valence $M = 4$}

We start by analysing the behaviour of monodisperse $M = 4$ patchy particles in a range of temperatures in which the system goes from a purely diffusive behaviour down to values of $T$ at which caging behaviour extending for almost three orders of magnitude in time is observed (see~\ref{app:msd}).

\begin{figure}[ht!]
   \centering
   \includegraphics[width=0.45\textwidth]{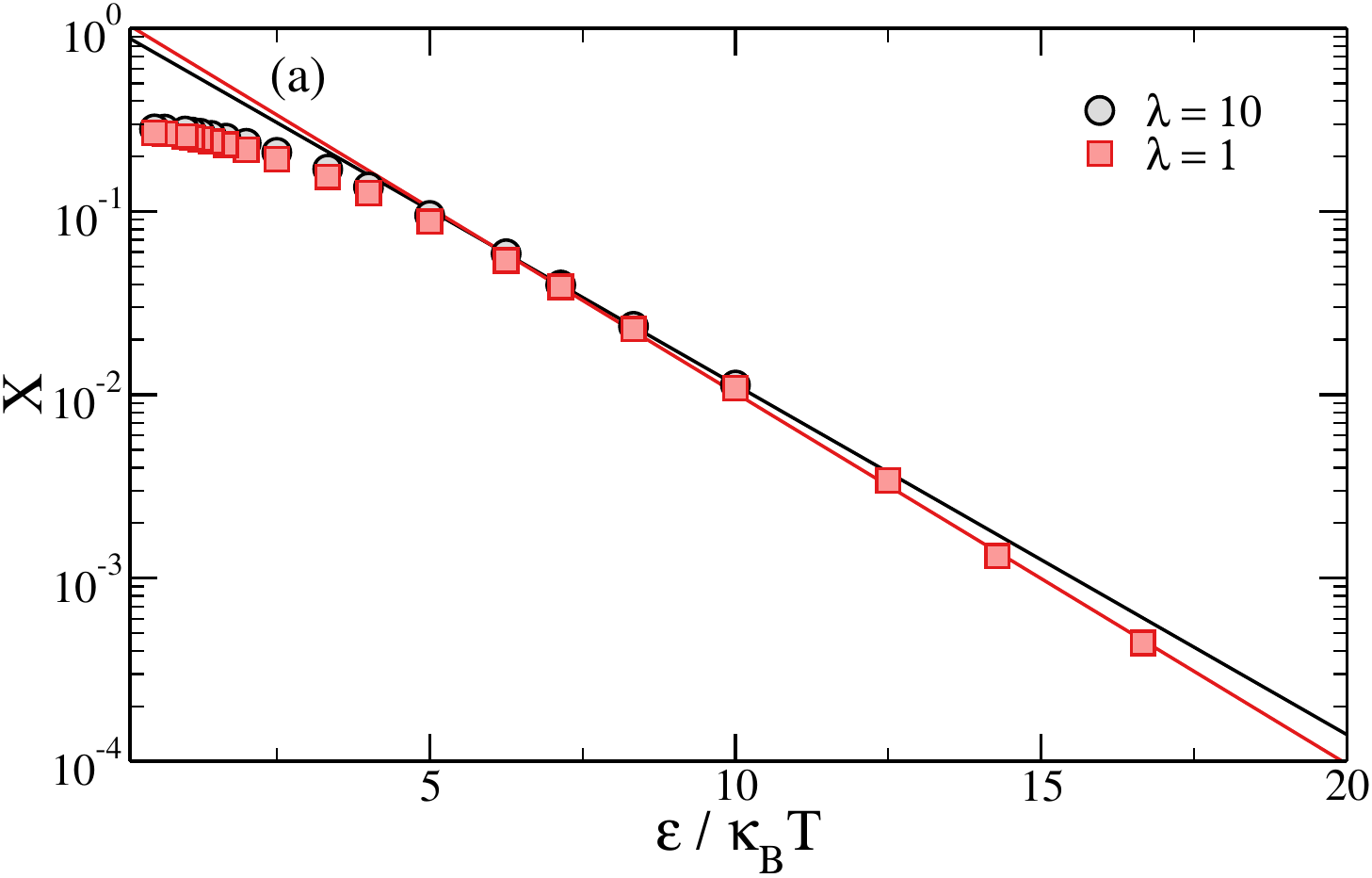} 
   \includegraphics[width=0.46\textwidth]{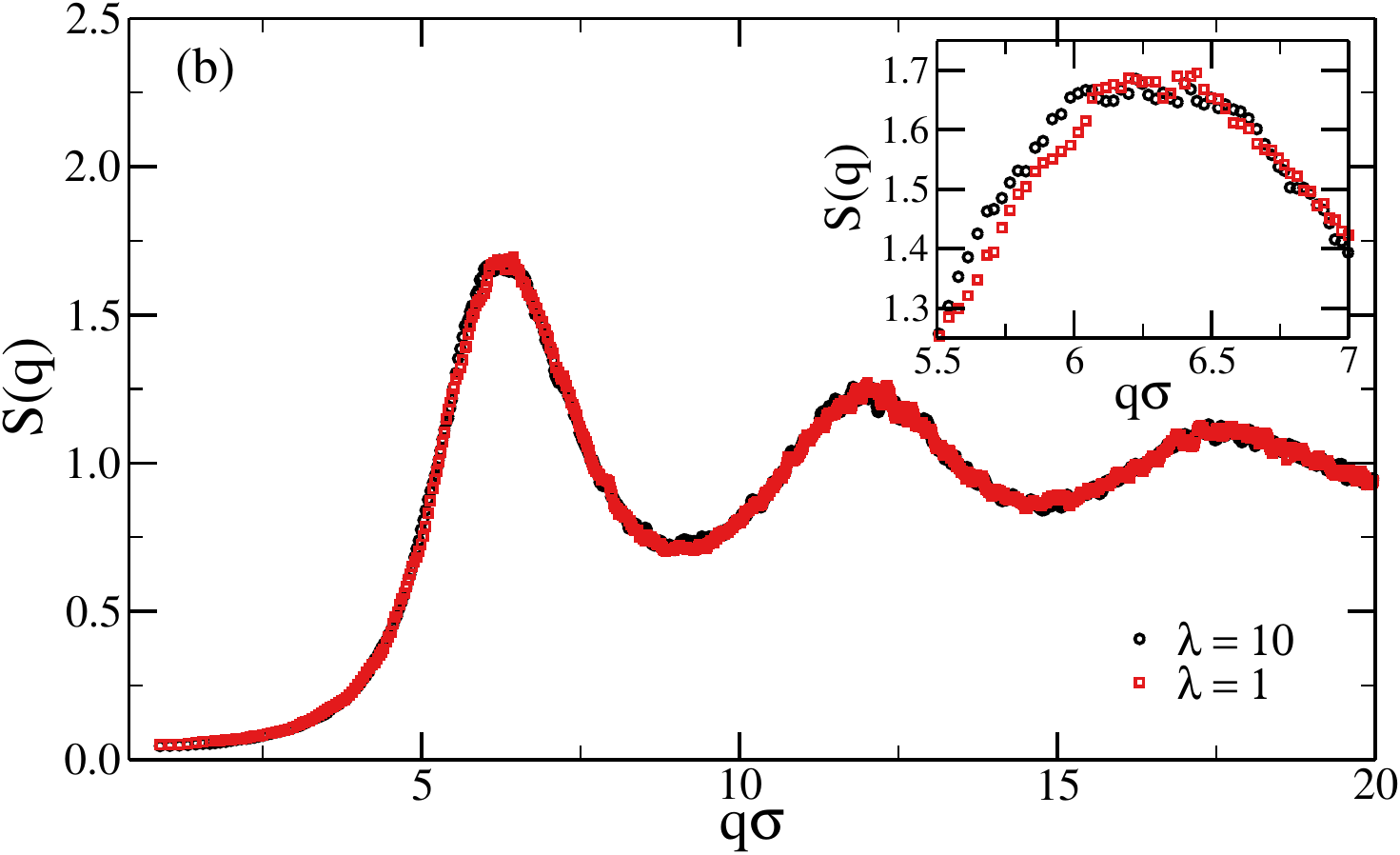} 
   \caption{(a) The probability that a patch is not involved in a bond, $X$, for a system of valence $M = 4$ at density $\rho \sigma^3 = 0.60$ as a function of inverse temperature with ($\lambda = 1$) and without ($\lambda = 10$) bond swapping. Symbols are simulation data, lines are fits to Arrhenius functions for the low temperature ($\varepsilon / k_B T > 6.5$) data. The activation energies are $-0.44\varepsilon = -0.48 \varepsilon_{\rm eff}$ for $\lambda = 10$ and $-0.47\varepsilon = -0.52\varepsilon_{\rm eff}$ for $\lambda = 1$ ($-0.45\epsilon$ if we consider the same range used for the $\lambda = 10$ data, \textit{i.e.} $6.5 < \varepsilon k_B T \leq 10$). (b) The structure factor of the same system simulated at temperature $k_B T / \varepsilon = 0.10$, with and without bond swapping. Inset: a zoom-in on the first peak.
}
\label{fig:M4_pb}
\end{figure}

Figure~\ref{fig:M4_pb}(a) shows the probability that a patch is not bonded, $X$, as a function of the inverse temperature at fixed density, $\rho \sigma^3 = 0.60$, with and without the bond swapping mechanism. At the lowest temperatures considered, $X \approx 0.01$ for $\lambda = 1$ and $X < 0.001$ for $\lambda = 10$, and in both cases there is an extended temperature range ($\varepsilon / k_B T > 6.5$) $X$ in which $X$ exhibits an Arrhenius behaviour with rather similar activation energies , $E_X = -0.44 \varepsilon$ for $\lambda = 10$ and $E_X = -0.46 \varepsilon$ for $\lambda = 1$. Both these values are rather close to the theoretical estimate of $0.5 \varepsilon_{\rm eff} = 0.455\varepsilon$~\cite{smallenburg2013liquids}, which assumes that the probability that a patch is bonded is independent of the state (bonded/unbonded) of the other patches belonging to the same particle.

We observe a tiny vertical shift between the two curves that is independent of temperature: This is caused by the fact that triplets of patches can freely form when bond-swapping is activated, while for $\lambda = 10$ triplet formation is energetically penalised. As a result, the overall volume available to the patches is larger in the former case, and therefore the effective density is smaller, leading to a slightly less bonded system. 

Figure~\ref{fig:M4_pb}(b) shows a comparison between the structure factors of a low-temperature system ($k_B T / \varepsilon = 0.10$) simulated with and without bond swapping. The two curves overlap almost perfectly, confirming that the structure of the systems is nearly independent of $\lambda$. Interestingly, a close look at the first-neighbour peak (located at $q \sigma \approx 2 \pi$, see inset) reveals a tiny broadening in the $\lambda = 10$ data towards low-wave vectors that is another signature of the slightly enhanced particle-particle repulsion due to the penalisation of triplet formation. We note on passing that no tetrahedral peak is present: its absence, due to the large size of the patches~\cite{saika2013understanding}, confirms the independent nature of the bonding state of each patch. Moreover, the absence of a network peak suggests that the thermodynamic equilibrium state is always disordered rather than crystalline even in the limit of large patch-patch attraction strengths~\cite{saika2013understanding,smallenburg2013liquids}, as observed in valence-limited systems with flexible bonds such as DNA nanostars~\cite{doi:10.1021/nn501138w}. Finally, we note that the structure factor is an increasing function of $q$ in the low-$q$ limit, signalling the near homogeneity of the system down to the largest probed length-scale.

\begin{figure}[ht!]
   \centering
   \includegraphics[width=0.45\textwidth]{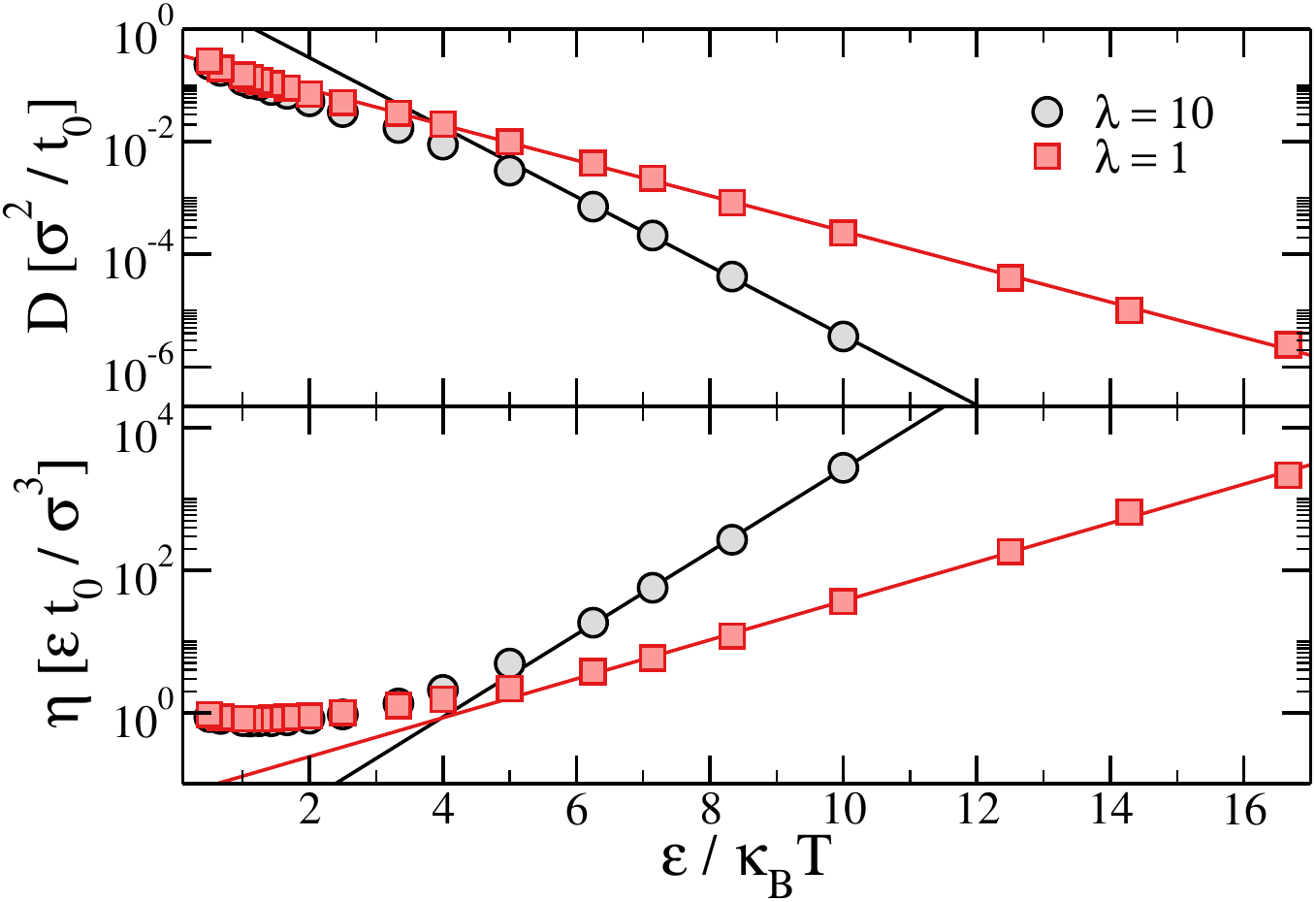} 
   \caption{Diffusion coefficient (top) and viscosity (bottom) for a system of valence $M = 4$ at density $\rho \sigma^3 = 0.60$ as a function of inverse temperature with ($\lambda = 1$) and without ($\lambda = 10$) bond swapping. Symbols are simulation data, lines are fits to Arrhenius functions for the low temperature ($\varepsilon / k_B T > 6.5$) data. For $D$, the activation energies are $-1.42 \varepsilon = 1.56 \varepsilon_{\rm eff}$ for $\lambda = 10$ and 
$-0.72\varepsilon = -0.79 \varepsilon_{\rm eff}$ for $\lambda = 1$. For $\eta$, the activation energies are $1.34\varepsilon = 1.47\varepsilon_{\rm eff}$ for $\lambda = 10$ and $0.63\varepsilon = 0.69 \varepsilon_{\rm eff}$ for $\lambda = 1$.}
\label{fig:M4_D_eta}
\end{figure}

Figure~\ref{fig:M4_D_eta} shows the diffusion constant and the viscosity as a function of inverse temperature. First of all, we note that the viscosity displays a minimum around $\varepsilon / k_B T = 1$, which hints at the presence of a crossover between liquid-like and gas-like behaviours~\cite{trachenko2020minimal}.
Focussing on the low temperature regime, we see that both quantities display an Arrhenius behaviours regardless of $\lambda$, and for $\lambda = 10$ the absolute value of the activation energy for $D$ and $\eta$ is similar and close to $\approx 1.5 \, \varepsilon_{\rm eff}$. Interestingly, experiments on DNA nanostars of different valence investigated under various conditions exhibit Arrhenius dynamics with comparable activation energies (ranging from $\approx 0.9$~\cite{conrad2019increasing} to $\approx 1.3$~\cite{C4SM02144D} and $\approx 2.3$~\cite{C8SM00751A} times the single-bond enthalpy).

Finally, we note that for this specific value of the valence we find that decreasing $\lambda$ from $10$ to $1$, \textit{i.e.} enabling bond-swapping, the activation energies for both $D$ and $\eta$ decrease by a factor that is close to two: in the investigated range of temperatures the swapping mechanism does not change the Arrhenius nature of diffusion and viscosity, but renormalise their effective activation energy.

\begin{figure}[ht!]
   \centering
   \includegraphics[width=0.45\textwidth]{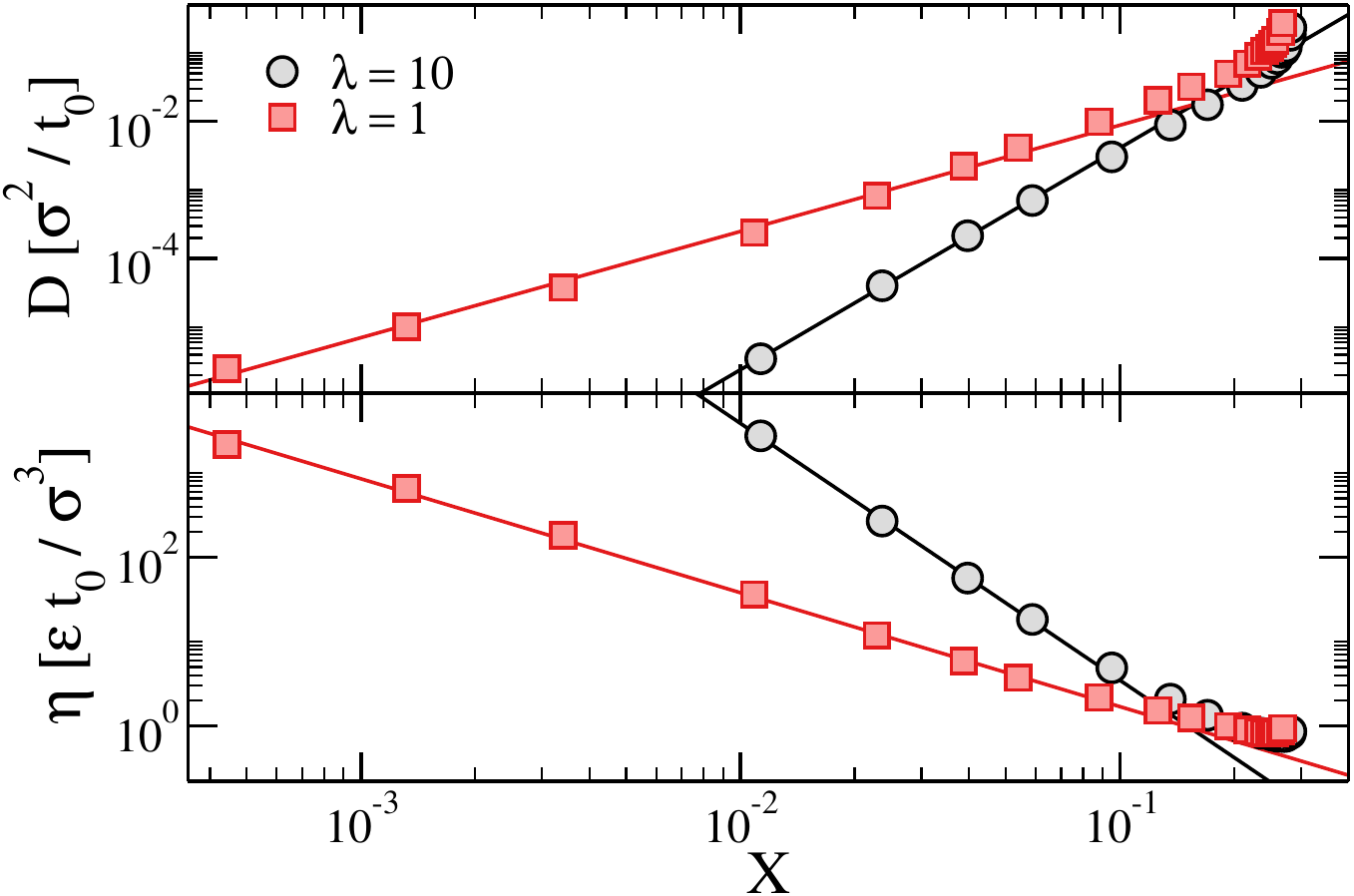} 
   \caption{Same data as Figure~\ref{fig:M4_D_eta}, but shown as a function of $X = 1 - p_b$ with a log-log scale. Symbols are simulation data, lines are fits to power-laws. For $D$, the fitted exponents are $\alpha_D = 3.23$ for $\lambda = 10$ and $\alpha_D = 1.55$ for $\lambda = 1$. For $\eta$, the fitted exponents are $\alpha_\eta = -3.04$ for $\lambda = 10$ and $\alpha_\eta = -1.35$ for $\lambda = 1$.}
\label{fig:M4_D_eta_X}
\end{figure}

To elucidate the role that bonding has on the dynamics we can also express $D$ and $\eta$ as a function of $X$. The parametric curves shown in Figure~\ref{fig:M4_D_eta_X} suggests that both $D$ and $\eta$ depend on $X$ as $X^\alpha$ in the same regime where all these quantities display an Arrhenius behaviour (\textit{i.e.} for small values of $X$). This feature signals that, even at this intermediate value of density, bonding is the main mechanism controlling the dynamics. 

Focussing on the system without swap, the exponent found for $D$ is $\alpha_D = 3.23$, which suggests that the activation energy is $3.23$ times the patch-patch bonding energy or, equivalently, that the diffusion process requires slightly more than three broken bonds. This is at odds with have been found with other tetravalent models, where a relationship $D \propto X^4$, which seems to suggest that diffusion requires four broken bonds, has been observed~\cite{doi:10.1021/jp056380y,doi:10.1021/la063036z,C6SM02282K}. However, all those foregoing models have a rather rigid tetrahedral geometry, as demonstrated by the presence of a tetrahedral peak in the structure factor~\cite{doi:10.1021/jp056380y,doi:10.1021/la063036z,hsu2008hierarchies,C6SM02282K}. By contrast, the model considered here is much more flexible, and it is perhaps this flexibility that makes diffusion require, on average, fewer broken bonds compared to more rigid models. This effect has been suggested in Ref.~\cite{smallenburg2013liquids}, where it was shown that, for a bond-swapping system, making patch-patch bonds more flexible weakened the dependence of $D$ on $\varepsilon / K_BT$.

Regarding systems with bond swapping ($\lambda = 1$), we see that the dynamics is also activated, albeit with a smaller activation energy, and therefore still dominated by the bonding process. Such an Arrhenius nature of the temperature-dependence of the diffusion constant can be understood by noting that a bond swap requires an unbonded patch, and the probability that a patch is unbonded depends exponentially on temperature. In particular, here the diffusion of bond-swapping particles seems to require approximately 1.5 broken bonds, roughly half of value observed for the $\lambda = 10$ case. This behaviour is reminiscent of what was found in the different but related limited-valence system studied in Ref.~\cite{doi:10.1021/acs.macromol.7b02186}: there, a vitrimer-like mixture of divalent and tetravalent particles were simulated in the $T \to 0$ limit, where the unbonded patches were provided by an unbalanced stochiometry, and $D$ was found to depend almost linearly on the fraction of unbonded patches~\cite{doi:10.1021/acs.macromol.7b02186}. The small difference between the two behaviours can perhaps be ascribed to the thermally-activated unbonding of the patches that acts here, but it is absent in the other model system.

\begin{figure}[ht!]
   \centering
   \includegraphics[width=0.45\textwidth]{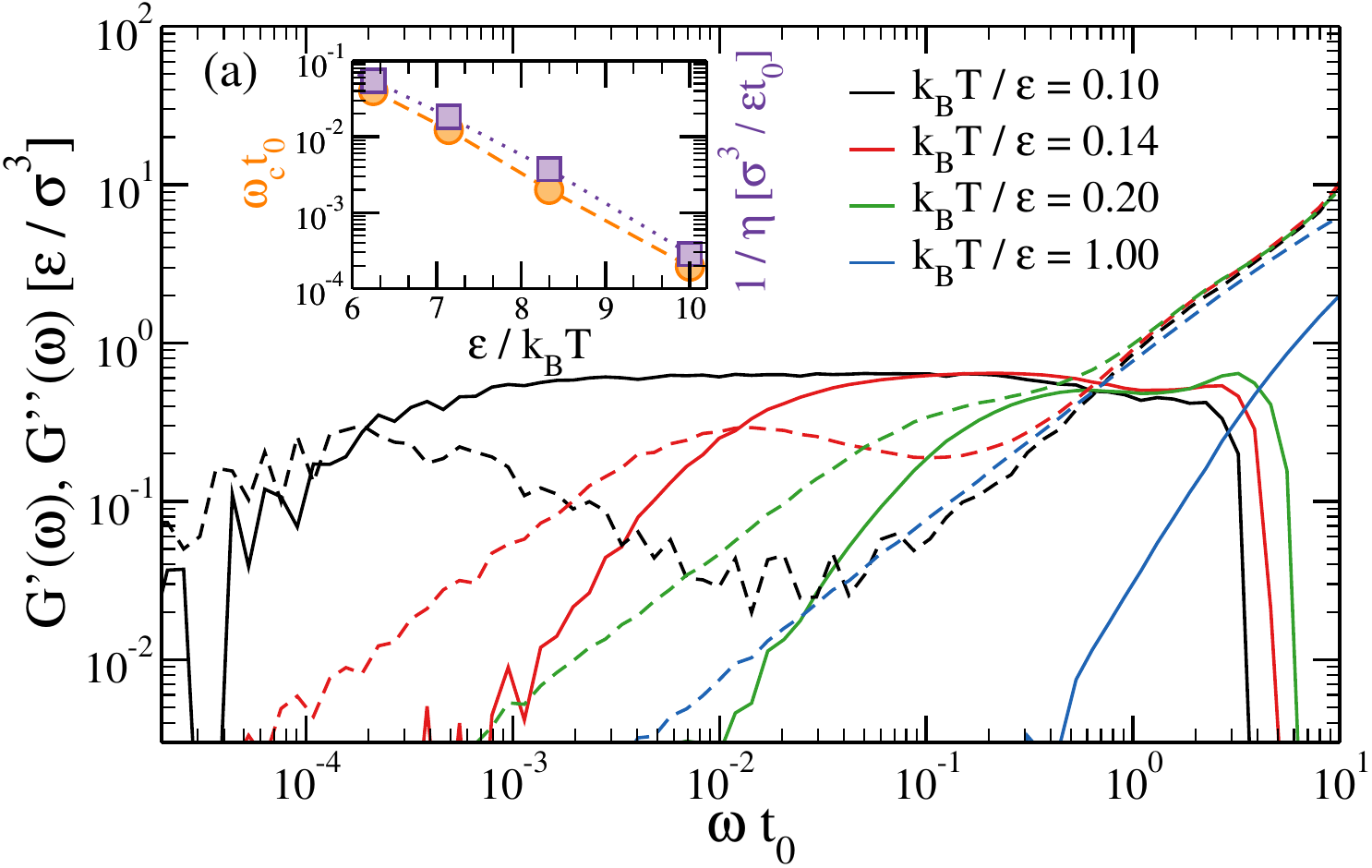} 
   \includegraphics[width=0.45\textwidth]{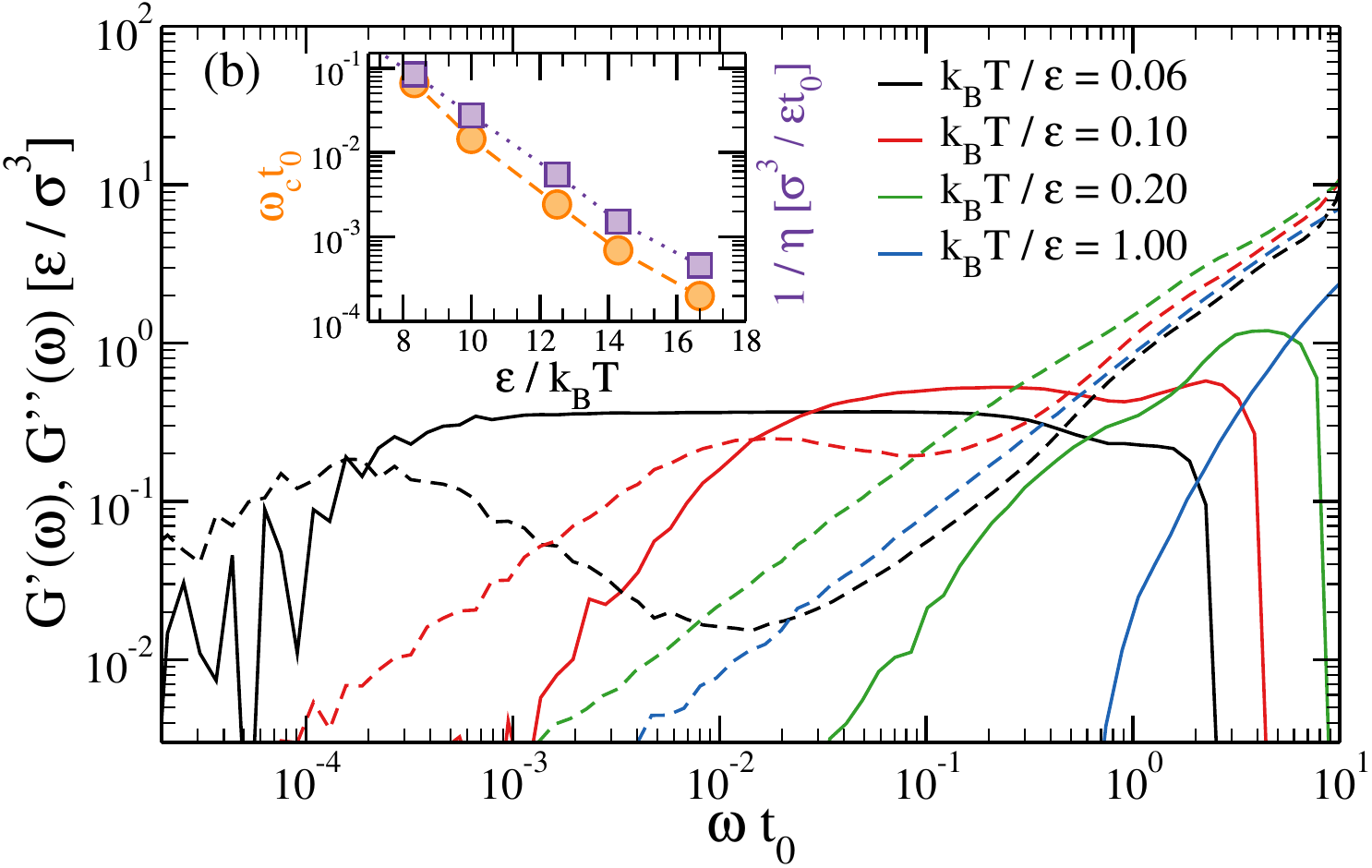} 
   \caption{Storage (full lines) and loss (dashed lines) moduli as a function of frequency for selected temperatures for systems of valence $M = 4$, density $\rho \sigma^3 = 0.60$ and (a) $\lambda = 10$ and (b) $\lambda = 1$. Insets: a comparison between the frequency at which $G'(\omega)$ and $G''(\omega)$ cross (orange circles, left axis) and the inverse of the viscosity (violet squares, right axis) as a function of inverse temperature.}
\label{fig:G_omega_M4}
\end{figure}

Next, we analyse the frequency response of the system by computing the storage and loss moduli $G'(\omega)$ and $G''(\omega)$. Results for selected cases spanning the investigated temperature range are shown in Figure~\ref{fig:G_omega_M4}. First of all, we note that the high-frequency ($\omega t_0 > 1$) response is linked to intra-bond vibrations that are model-dependent and therefore are not analysed here.

Secondly, in all cases $G'(\omega) \sim \omega^2$ and $G''(\omega) \sim \omega$ in the $\omega \to 0$ limit, signalling that all systems behave like liquids at sufficiently long times. While at high and intermediate temperatures the systems have a response that is essentially viscous (\textit{i.e.} $G''(\omega)$ is always larger than $G'(\omega)$), upon lowering $T$ the storage modulus becomes larger than the loss modulus for a range of frequencies, and exhibits a plateau that signals the presence of a solid-like (elastic) response. The extent of the plateau (in frequency) roughly corresponds to the width of the plateau observed in the mean-squared displacement, shown in  Fig.~\ref{fig:M4_msd}, which in turn corresponds to the regime in which particles wiggle in the cage formed by their bonded neighbours, as if they were vibrating around their equilibrium positions in a solid. However, as soon as a particle breaks or swaps enough bonds, it restructures its cage and can diffuse, thus contributing to the viscous relaxation. The inverse of the time at which this happens can be estimated by looking at the crossing frequency of the storage and loss moduli, $\omega_c$, and it is a measure of the final relaxation time of the system. Indeed, as shown in the insets of Fig.~\ref{fig:G_omega_M4}, $\omega_c$ displays a temperature dependence that closely follows that of the viscosity.

The picture described above is shared by the swapping and non-swapping systems, whose behaviours qualitatively differ only for what concerns the height of the elastic plateau. Indeed, while the $\lambda = 10$ system displays a rather temperature-independent plateau, in the swapping system the height of the plateau decreases by about $\approx 30\%$ going from $k_B T / \varepsilon = 0.12$ down to $k_B T / \varepsilon = 0.06$. The additional channel of relaxation provided by the bond swapping, which becomes relatively more important as $T$ decreases, makes it possible to partially relax the stress without the need of breaking any bonds, thus effectively lowering the elastic response of the system.

\section{The effect of the valence}

Here we show results of binary mixtures composed of divalent and $k$-valent particles, with $k = 3, 4$. The compositions investigated are reported in Table~\ref{tbl:mixtures}.

The qualitative behaviour of the structure factor, bonding probability $p_b$ (and hence of $X$), of the viscosity and of the diffusion constant is shared by all the systems investigated (pure $M=2$ and $M=3$ systems, as well as mixtures of $M=2, 3$ and $M=2, 4$ particles): at low temperature $S(q) \to 0$ for $q \to 0$, $X$, $\eta$ and $D$ all display an Arrhenius dependence on $T$, and the viscosity and diffusion constant depend on $X$ as power laws that directly connects these quantities to the fraction of broken bonds. This behaviour on one hand demonstrates the prevalence of bonding in dictating the properties of these limited-valence systems, and on the other hand suggests that the dynamics of the binary mixtures investigated here can be understood in terms of a pure system composed of particles having an effective valence.

\begin{figure}[ht!]
   \centering
   \includegraphics[width=0.45\textwidth]{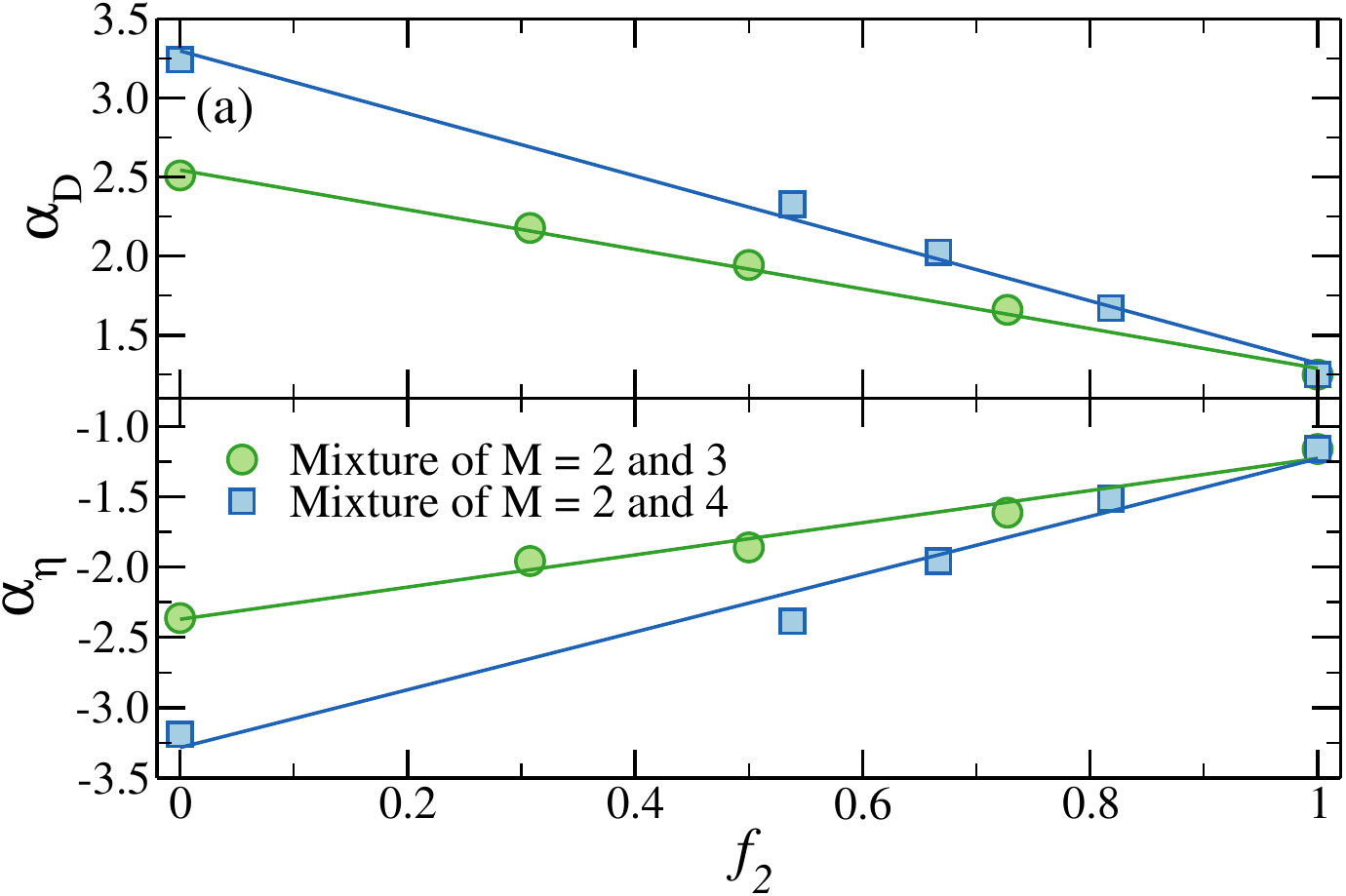} 
   \includegraphics[width=0.45\textwidth]{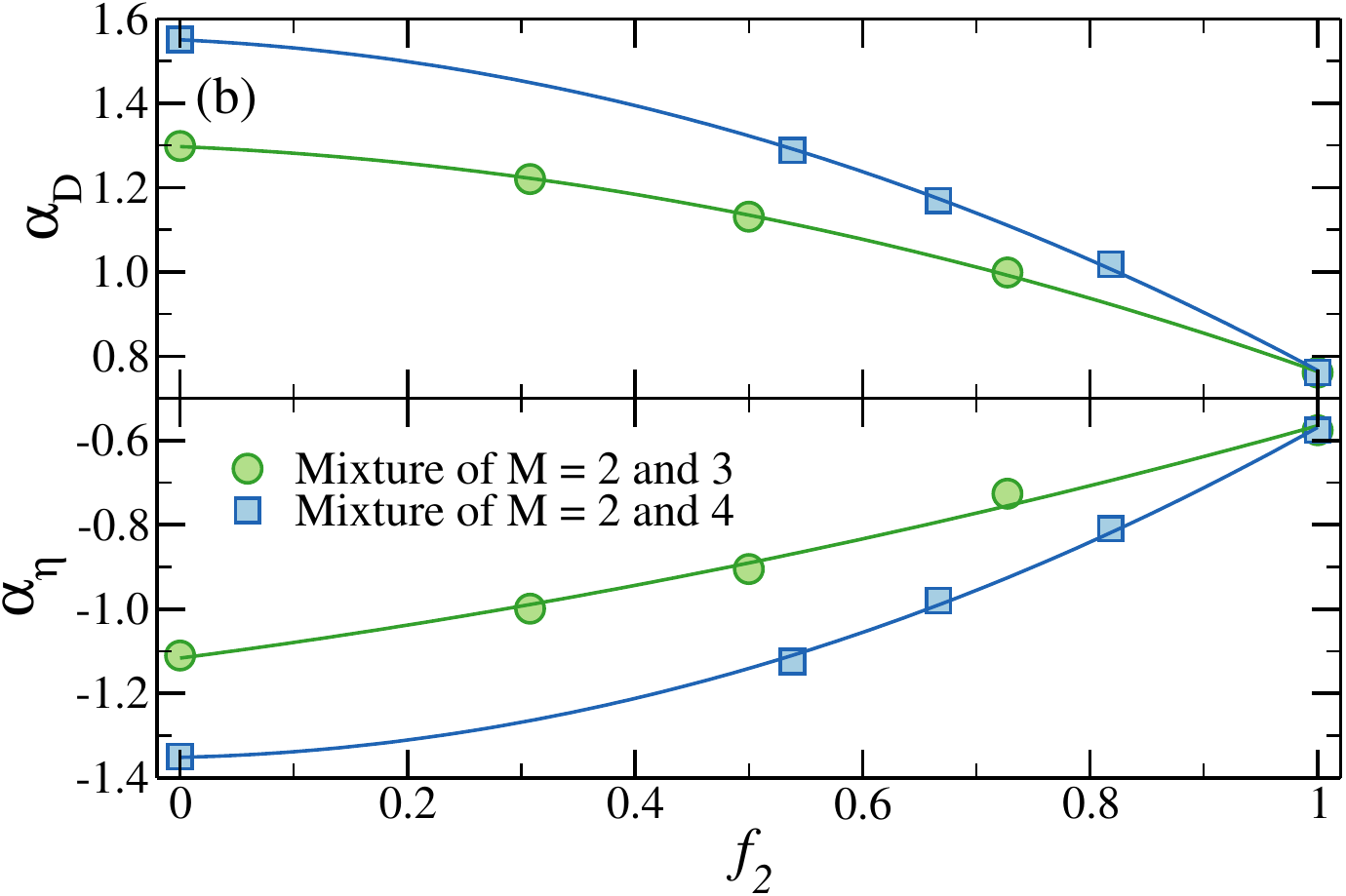} 
   \caption{Symbols are power-law exponent connecting the diffusion constant (up) and viscosity (bottom) of low-temperature patchy systems with different fractions of $M = 2$ particles mixed with $M = 3$ or $M = 4$ particles at $\rho\sigma^3 = 0.60$, as a function of the fraction of divalent patches for (a) $\lambda = 10$ and (b) $\lambda = 1$. Solid lines are (a) linear and (b) quadratic best fits.}
\label{fig:mixtures_eta}
\end{figure}

Figure~\ref{fig:mixtures_eta} shows the $\alpha_D$ and $\alpha_\eta$ exponents connecting $D$ and $\eta$ with $X$ for all considered systems, with and without the swap, as a function of the fraction of divalent patches $f_2$, defined in Eq.~\eqref{eq:f2}. Interestingly, when plotted as a function of $f_2$, the exponents of the diffusion constant and of the viscosity of the $\lambda = 10$ binary mixtures interpolate linearly between the values observed in the pure systems, suggesting that there exists a simple relationship between the average valence of a binary mixture and the effective valence of its associated one-component description.

By contrast, if the bond-swapping mechanism is active, \textit{i.e.} for $\lambda = 1$, the exponents take smaller absolute values compared to the corresponding $\lambda = 1$ cases, and their dependence on $f_2$ is markedly non-linear. Compared with the $\lambda = 10$ case, where the dynamics is fully enslaved to bond breaking, here there is an additional decorrelation channel due to the swap. Although, as mentioned earlier, the swapping mechanism is also thermally activated, since it requires unbonded patches, its dependence on the fraction of broken bonds is different from the one observed when only the bond-breaking mechanism is in place (see, \textit{e.g.}, Ref.~\cite{doi:10.1021/acs.macromol.7b02186} for a specific example). As demonstrated by the solid lines in Figure~\ref{fig:mixtures_eta}(b), the observed behaviour is compatible with a (phenomenological) quadratic dependence on $f_2$.

\subsection{Stokes-Einstein relation}

The diffusion constant $D$ of a tracer of hydrodynamic diameter $\sigma_H$ suspended in a simple liquid of viscosity $\eta$ is well-described by the Stokes-Einstein (SE) relation, which can be written as~\cite{sutherland1905lxxv,einstein1905molekularkinetischen}

\begin{equation}
\label{eq:SE}
D = C \frac{k_B T}{\sigma_H \eta },
\end{equation}

\noindent where $C$ is a constant that depends on the geometry and surface (boundary) properties of the solute. 

Although Eq.~\eqref{eq:SE} is usually derived by considering the solvent as a continuous medium, it often applies to molecules belonging to the solvent itself~\cite{hansen2013theory}. In these cases, the $\sigma_H$ term appearing in Eq.~\eqref{eq:SE} has been challenged (see \textit{e.g.} Refs.~\cite{ohtori2018stokes,costigliola2019revisiting,khrapak2021excess}) by leveraging expressions derived independently by Eyring and Ree~\cite{eyring1961significant} and Zwanzig~\cite{zwanzig1983relation}, where $\sigma_H$ is replaced by $\rho^{-1/3}$:

\begin{equation}
D = \alpha_{SE} \frac{k_B T \rho^{1/3}}{\eta},
\end{equation}

\noindent which implies

\begin{equation}
\label{eq:SE_rho}
\frac{D}{\rho^{1/3}} = \alpha_{SE} \frac{k_B T}{\eta}.
\end{equation}

In the following we will use Eq.~\eqref{eq:SE_rho} rather than Eq.~\eqref{eq:SE}. According to Zwanzig's theory, 

\begin{equation}
\alpha_{SE} = \frac{D\eta}{k_B T\rho^{1/3}} \approx 0.132 (1 + \eta / 2\eta_l),
\label{eq:alpha}
\end{equation}

\noindent
where $\eta_l$ is the longitudinal viscosity, and therefore $\alpha_{SE}$ can only take values between 0.132 and 0.181~\cite{zwanzig1983relation}. However, Zwanzig's theoretical approach rests on several assumptions, the main one being that the liquid under investigation exhibits solid-like oscillations, and it is therefore expected to work well only close to the liquid-solid phase transition.

The SE relation is known to break down in complex systems (\textit{e.g.} supercooled~\cite{becker2006fractional} and ionic~\cite{harris2009fractional} liquids). It has been noted that often in these cases the relationship between $D$, $\eta$ and $T$ can be described by a fractional SE relation (FSE)~\cite{becker2006fractional,harris2009fractional,costigliola2019revisiting}, \textit{viz.}

\begin{equation}
\label{eq:FSE}
D \propto \left( \frac{T^*}{\eta^*} \right)^s,
\end{equation}

\noindent where the $^*$ subscript signals dimensionless units.

\begin{figure}[ht!]
   \centering
   \includegraphics[width=0.45\textwidth]{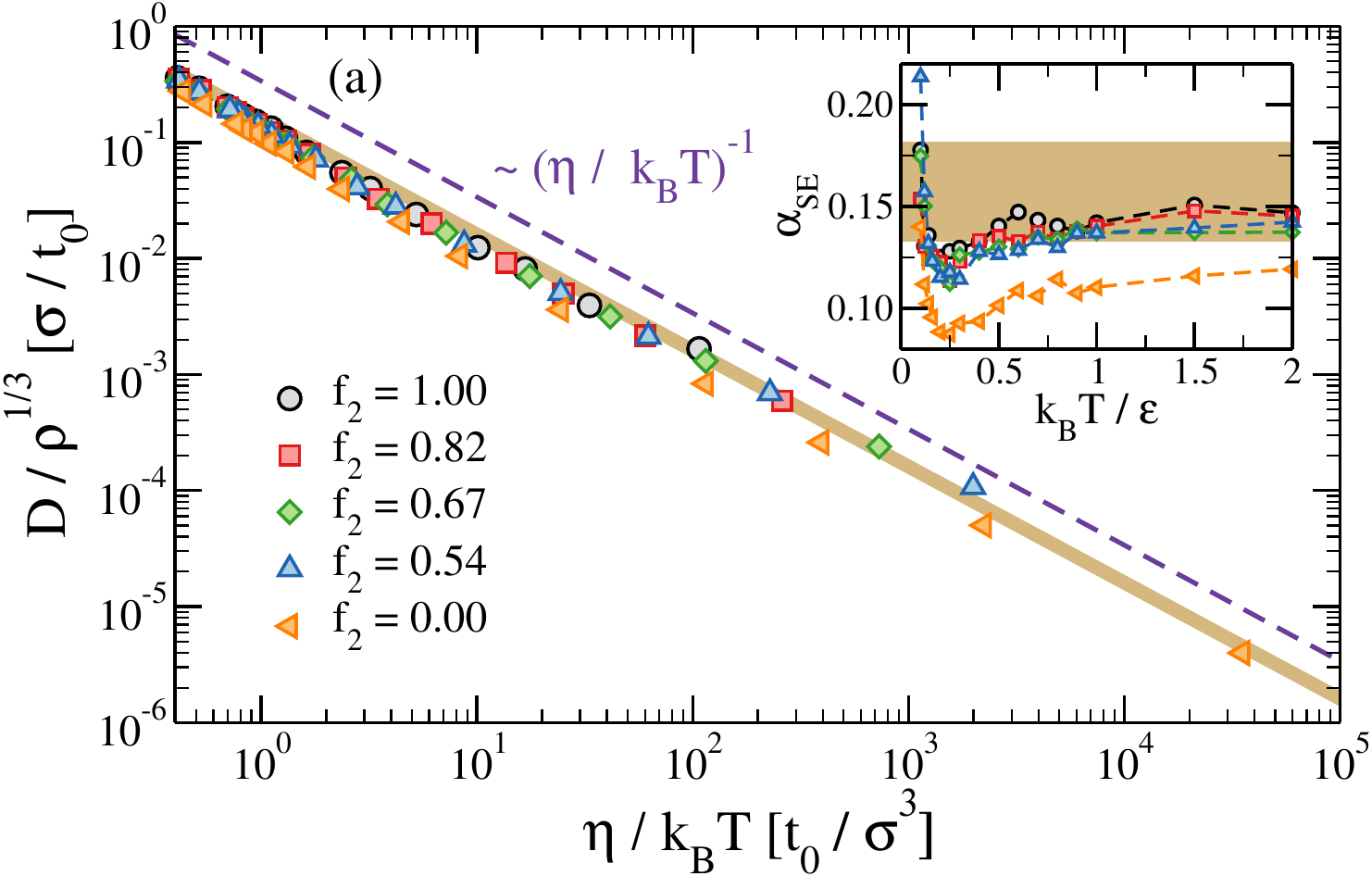} 
   \includegraphics[width=0.45\textwidth]{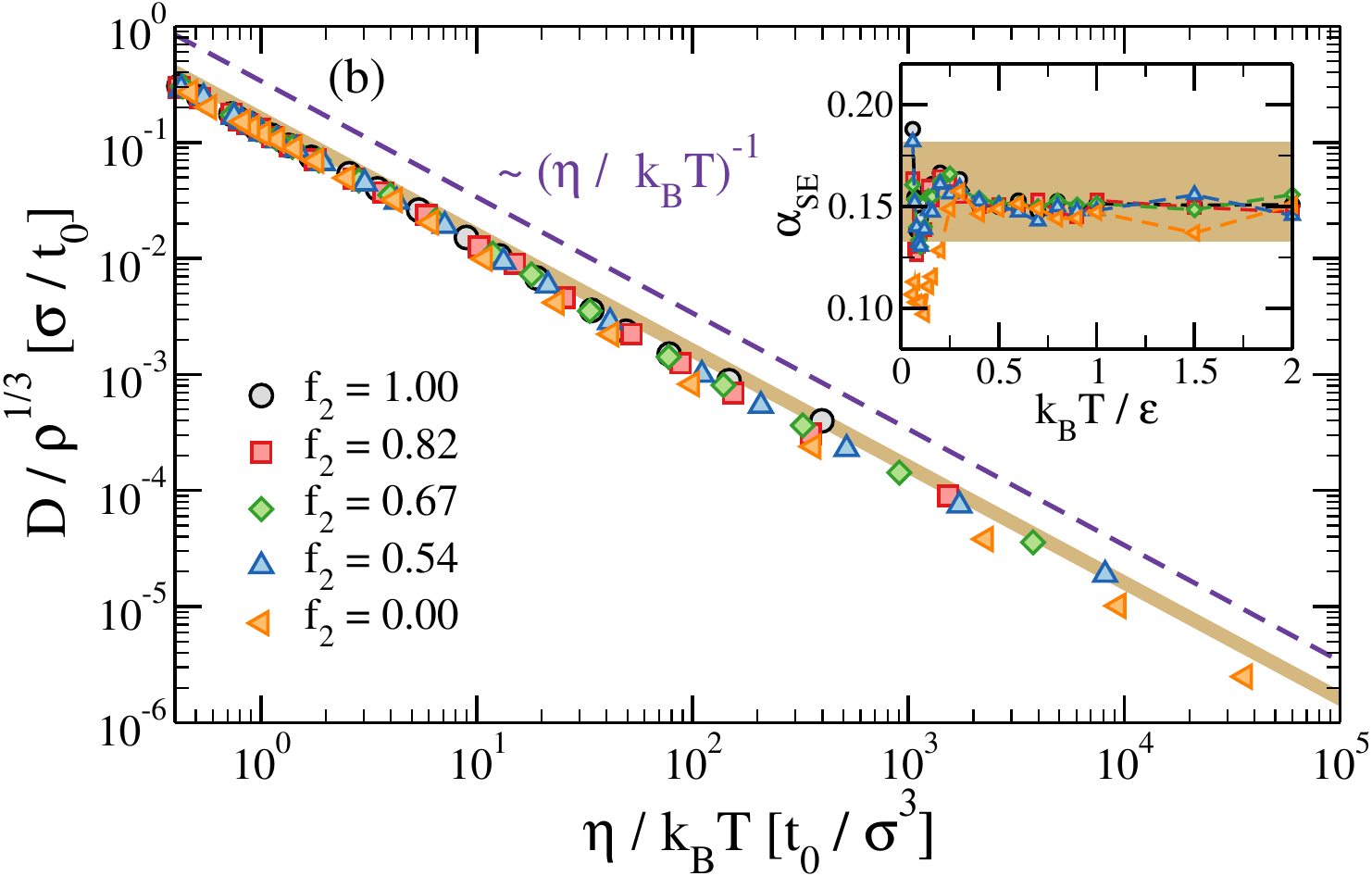} 
   \caption{Test of the Stokes-Einstein relation, Eq.~\eqref{eq:SE_rho}, for the (a) non-swapping ($\lambda = 10$) and (b) swapping ($\lambda = 1$) systems composed of $M = 4$ particles  only ($f_2 = 0$) or mixed with divalent particles ($f_2 > 0$), simulated at $\rho\sigma^3 = 0.60$. Here the diffusion constant $D$ is multiplied by the average interparticle distance, $\rho^{-1/3}$, and plotted as a function of $\eta / k_B T$ on a log-log scale. The violet line is the slope predicted by Eq.~\eqref{eq:SE}, while the shaded brown region bounds the theoretical range as predicted by Zwanzig's theory~\cite{zwanzig1983relation}. The insets show the numerical value of $\alpha_{SE}$, Eq.~\eqref{eq:alpha}, for the same systems.}
\label{fig:M4_SE}
\end{figure}

Figure~\ref{fig:M4_SE} shows that the dependence of the diffusion constant on the shear viscosity is compatible with the Stokes-Einstein relation without hydrodynamic diameter, Eq.~\eqref{eq:SE_rho}, for all the systems investigated (pure and binary mixtures, with and without swap) made of $M = 4$ and $M = 2$ particles. The data on pure systems and binary mixtures made of trivalent patchy particles exhibit the same qualitative features and are shown in~\ref{app:SE_M3}.

The data spans five orders of magnitude for both $D$ and $\eta$, and on the scale of the plot displays a rather good collapse. Interestingly, even at low temperature (\textit{i.e.} in the large-viscosity limit), when the systems behave like supercooled liquids exhibiting \textit{e.g.} caging effects (see Fig.~\ref{fig:M4_msd}), the inversely linear relationship between $D$ and $\eta$ holds. We thus do not see significant deviations from the non-fractional SE relation even under conditions where the systems we study behave like supercooled liquids, at odds with what has been observed with more realistic interaction potentials~\cite{becker2006fractional,xu2009appearance,pan2017structural}. This may possibly change at temperatures lower than those investigated here, since the value of $\alpha_{SE}$, shown in the insets, seems to steeply increase upon decreasing $T$, perhaps signalling a possible break down of the SE relation at very low temperatures.

Interestingly, many data points, mostly belonging to the non-swapping systems, lie below the expected range set by Zwanzig's theory~\cite{zwanzig1983relation}, as shown in the insets. This disagreement can be rationalised by considering that, contrary to the shear viscosity, the diffusion constant is known to depend on the system size~\cite{10.1063/1.465445,PhysRevE.68.021203,doi:10.1021/jp0477147}. Leveraging hydrodymamics arguments, it has been shown that, for incompressible fluids and at leading order in the (cubic) box size $L$, the relationship between $D = D(L)$ and its thermodynamic limit, $D_0 \equiv \lim_{L\to \infty} D(L)$ takes the form

\begin{equation}
D(L) = D_0 - \frac{k_B T \xi }{6 \pi \eta L},
\label{eq:D_0}
\end{equation}

\noindent
where $\xi \approx 2.837298$ is a geometrical constant, and $\eta$ does not exhibit considerable finite-size effects~\cite{10.1063/1.465445,doi:10.1021/jp0477147}. Eq.~\eqref{eq:D_0} has been shown to work remarkably well for many systems, ranging from Lennard-Jones to water, biomolecules, and complex mixtures~\cite{doi:10.1080/08927022.2020.1810685}. However, a detailed study of the diffusion constant of hard spheres has shown that packing fraction, or, equivalently, density, may play a role in determining the functional dependence of $D$ on the system size~\cite{doi:10.1021/jp067373s}.

\begin{figure}[ht!]
   \centering
   \includegraphics[width=0.45\textwidth]{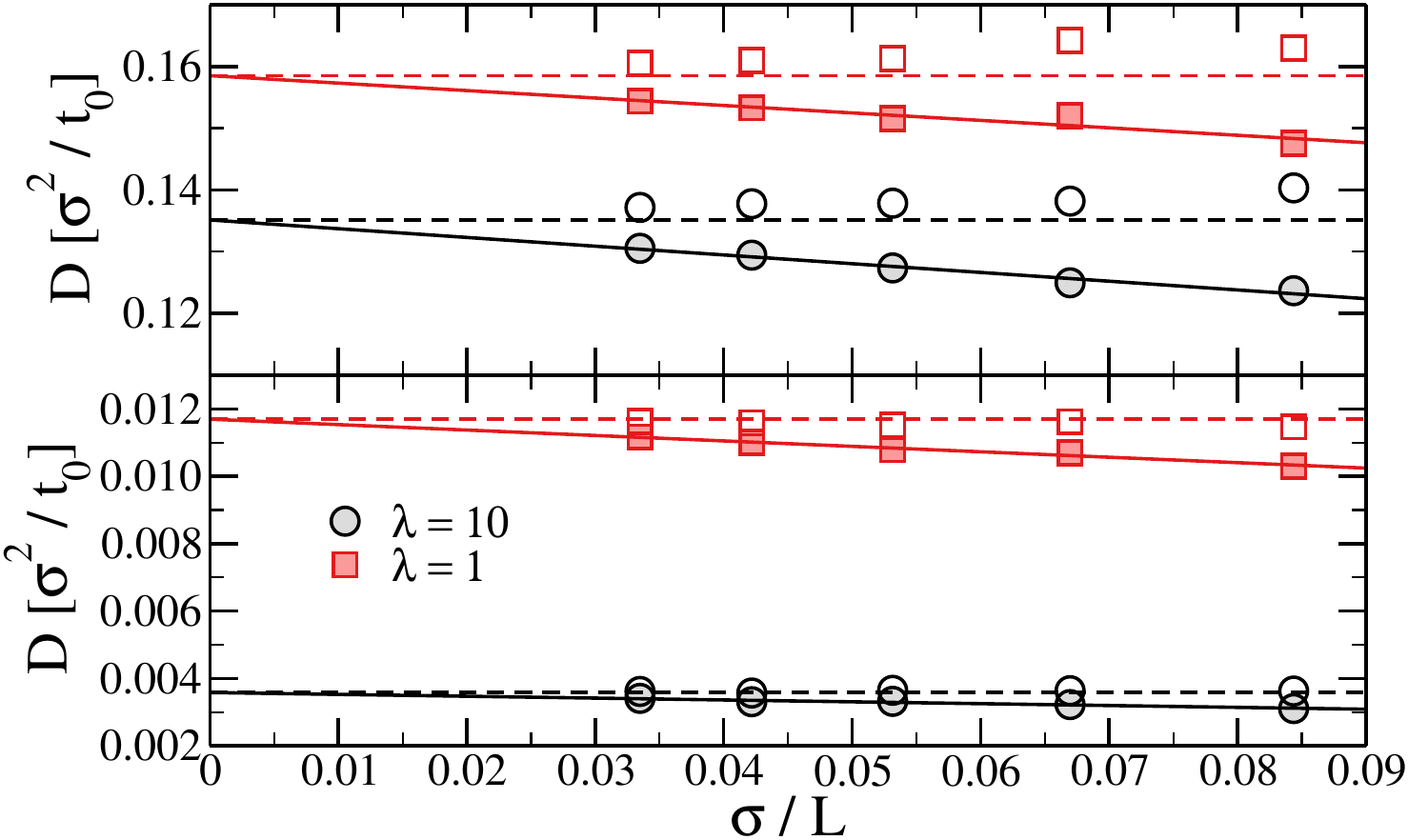}
   \caption{ The system size dependence of the diffusion coefficent of $M = 4$, $\rho\sigma^3 = 0.60$ systems for (top) $\varepsilon / k_B T = 1.0$ and (bottom) $\varepsilon / k_B T = 5.0$. Full symbols are simulation data obtained by simulating systems made of $N = 1000$, $2000$, $4000$, $8000$ and $16000$ particles at constant density. The full line is a linear fit to the function $D(L) = D_{\rm 0, fit} + A / L$, with $D_{0, \rm fit}$ and $A$ fitting parameters. The dashed line is the $D(L) = D_{\rm 0, fit}$ locus, while empty symbols are the values of $D_0$ obtained by applying Eq.~\eqref{eq:D_0} with the shear viscosity computed independently.}
\label{fig:D_inf_T}
\end{figure}

In Figure~\ref{fig:D_inf_T} we test the proposed scaling picture for the $M = 4$ system and two temperatures (high, $\varepsilon / k_B T = 1.0$, and low, $\varepsilon / k_B T = 5.0$) by fixing the density, $\rho\sigma^3 = 0.60$, and simulating different number of particles $N = 1000$, $2000$, $4000$, $8000$, and $16000$. We see that in all cases there is a system-size dependence that is always compatible with a linear relationship connecting $D(L)$ and $\eta$, and that the fitted thermodynamic limit of the diffusion constant $D_{0, \rm fit}$ is always $\approx 10\%$ larger than the value obtained in $N = 1000$ simulations. Interestingly, directly applying Eq.~\eqref{eq:D_0} by using $\eta$ to obtain $D_0$ from the finite-size $D(L)$ yields an agreement with the extrapolated values that somewhat depends on $T$: while at high temperature $D_0$ is always marginally larger than $D_{0, \rm fit}$ by a few percent, at low $T$ the agreement is almost perfect. Notwithstanding this small difference, these results show that the patchy particle model under study, at least at this intermediate density, can be added to the list of systems for which the D\"unweg-Kremer-Yeh-Hummer correction works. 

If we neglect the weak temperature-dependence of the finite-size correction, we can use Eq.~\eqref{eq:D_0} to estimate the infinite-size diffusion constants and use those in the SE relation.

\begin{figure}[ht!]
   \centering
   \includegraphics[width=0.45\textwidth]{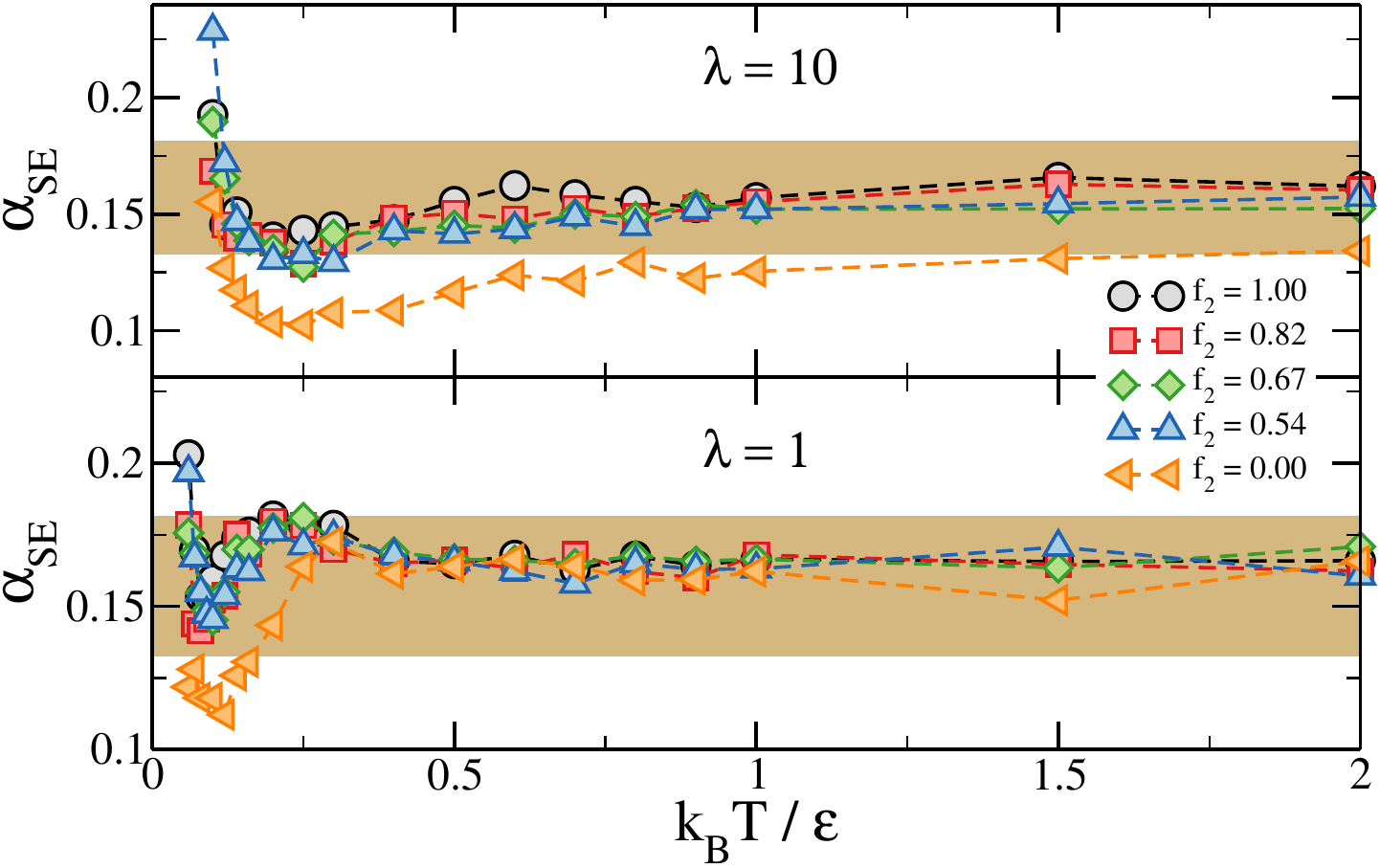}
   \caption{The SE constant $\alpha_{SE}$ computed by using $D_0$ instead of $D$ in Eq.~\eqref{eq:alpha} for the (top) non-swapping ($\lambda = 10$) and (bottom) swapping ($\lambda = 1$) systems composed of $M = 4$ particles only ($f_2 = 0$) or mixed with divalent particles ($f_2 > 0$), simulated at $\rho\sigma^3 = 0.60$. The shaded brown region bounds the theoretical range as predicted by Zwanzig's theory~\cite{zwanzig1983relation}.}
\label{fig:shifted_SE}
\end{figure}

Comparing Figure~\ref{fig:shifted_SE} with the insets of Figure~\ref{fig:M4_SE} it is clear that using $D_0$ instead of $D$ in Eq.~\eqref{eq:alpha} yields values of $\alpha_{SE}$ that are in better agreement with the range predicted by Zwanzig's theory~\cite{zwanzig1983relation}. However, deviations can still be observed, with $\alpha_{SE}$ taking values below what Eq.~\eqref{eq:alpha} would allow in some cases, \textit{e.g.}, the $f_2 = 0$ (\textit{i.e.} pure $M=4$), $\lambda = 10$ system at high and intermediate temperatures and the $f_2 = 0$, $\lambda = 1$ system at intermediate temperatures. Moreover, the tendency of all the curves to steeply increase upon decreasing temperature is still present, and confirms that a stronger breakdown of the SE relation may occur at very low $T$, in line with what has been observed in some supercooled liquids~\cite{jung2004excitation,becker2006fractional}. 

\subsection{Rheological properties}

We now analyse the effect of a change of composition on the frequency response of valence-limited fluids.

\begin{figure}[ht!]
   \centering
   \includegraphics[width=0.45\textwidth]{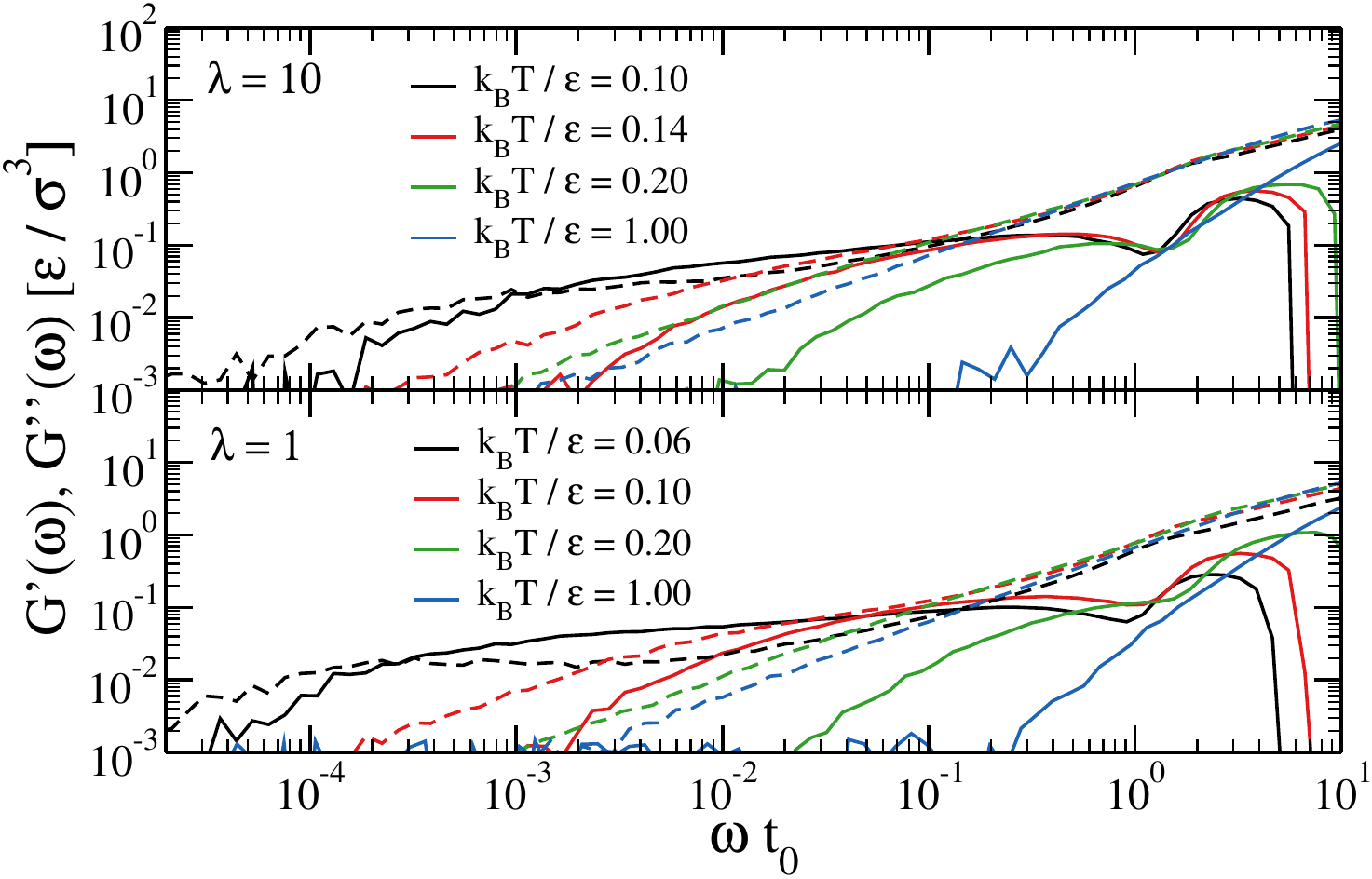} 
   \caption{Storage (full lines) and loss (dashed lines) moduli as a function of frequency for selected temperatures for systems composed of divalent and trivalent particles with $f_2 = 0.5$, simulated at density $\rho \sigma^3 = 0.60$ and (top) $\lambda = 10$ and (bottom) $\lambda = 1$.}
\label{fig:G_omega_M24}
\end{figure}

From the qualitative point of view, all investigated mixtures behave in the same way, which can be appreciated in Figure~\ref{fig:G_omega_M24}, where the storage and loss moduli for a mixture of divalent and trivalent particles with $f_2 = 0.5$ are shown for selected temperatures. In all cases we observe the expected terminal relaxation, \textit{e.g.} $G'(\omega) \sim \omega^2$ and $G''(\omega) \sim \omega$ for $\omega \to 0$. However, at odds with the results of pure $M = 4$ (and $M = 3$, see below) systems, $G'(\omega)$ never plateaus, even for those frequencies for which $G'(\omega) > G''(\omega)$. Moreover, at low enough temperature $G'(\omega)$ exhibits a functional dependence on $\omega$ that is compatible with a power law extending for $1-3$ decades, depending on the system.

Power-law rheology is a feature that has been observed in many complex materials that are characterised by multiple relaxation times, ranging from cells~\cite{PhysRevLett.87.148102}, to gluten~\cite{10.1122/1.2828018} and colloidal~\cite{10.1122/1.5025622} gels, hydrogels~\cite{doi:10.1021/acsmacrolett.5b00597}, complex interfaces~\cite{doi:10.1098/rspa.2012.0284}, and DNA nanostars~\cite{conrad2023towards}. The emerging power-law behaviours are often characterised in terms of the fractional Maxwell model~\cite{SCOTTBLAIR194721,10.1122/1.5025622,doi:10.1098/rspa.2012.0284}, or directly as a sum of exponential relaxation modes~\cite{doi:10.1021/ma00217a026,10.1122/1.2828018}.

\begin{figure}[ht!]
   \centering
   \includegraphics[width=0.45\textwidth]{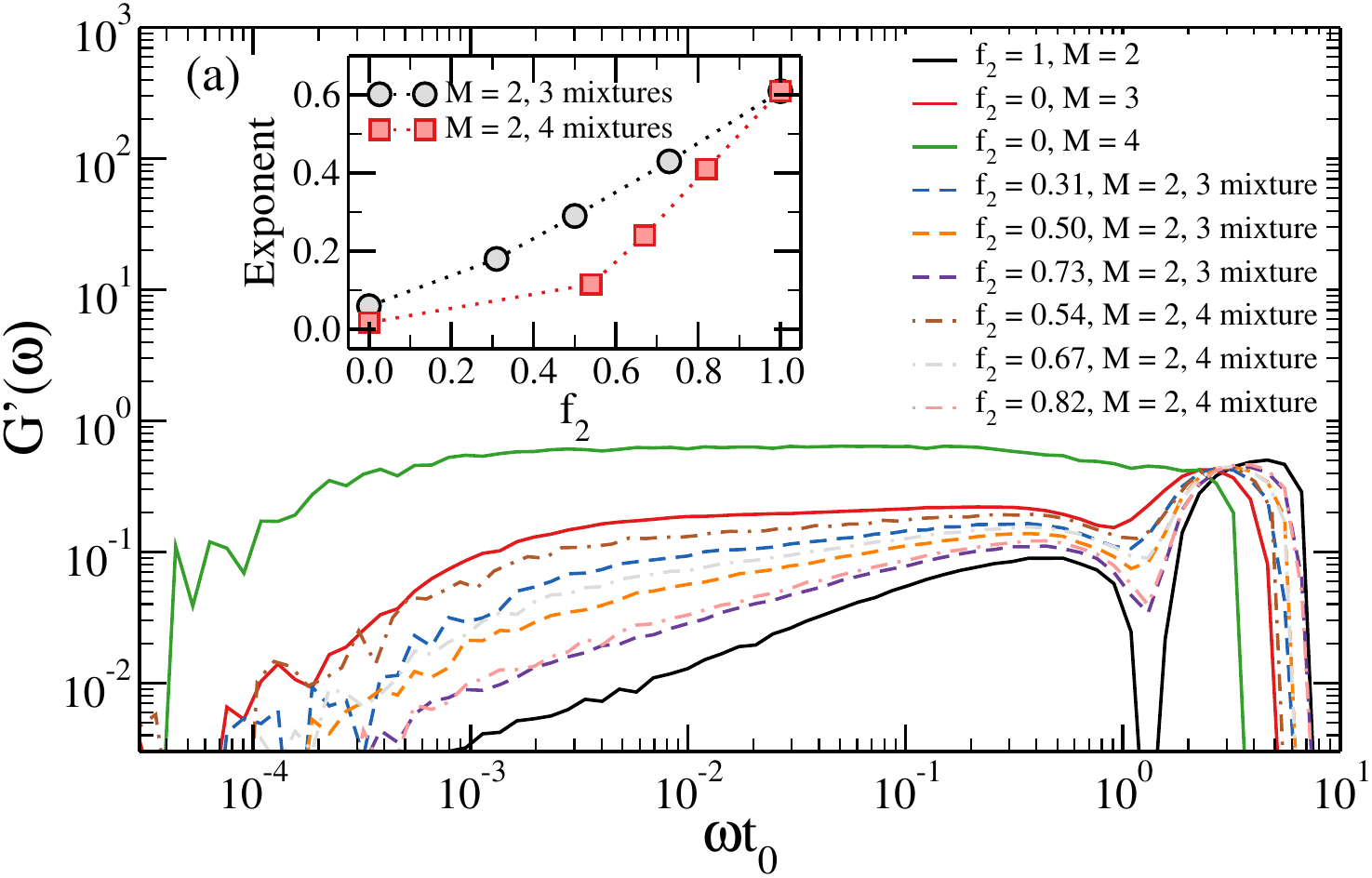} 
   \includegraphics[width=0.45\textwidth]{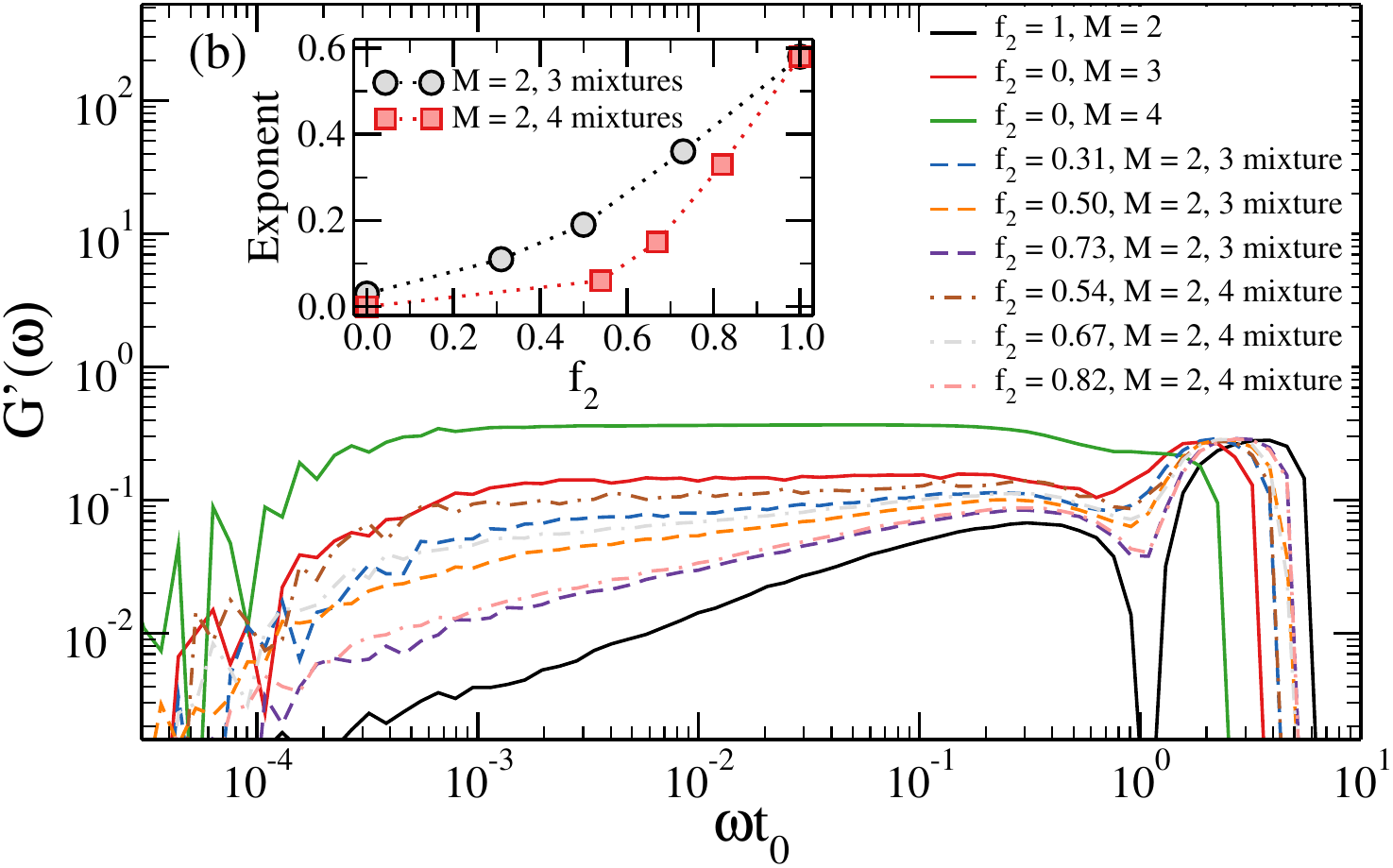} 
   \caption{Storage moduli as a function of frequency for all the investigated pure and mixture systems, simulated at density $\rho \sigma^3 = 0.60$ and (a) $\lambda = 10$ and (b) $\lambda = 1$ at the lowest temperatures investigated ($k_B T / \varepsilon = 0.10$ for $\lambda = 10$ and $k_B T / \varepsilon = 0.06$ for $\lambda = 1$). Full lines refer to pure systems, dashed lines to $M = 2, 3$ mixtures and dashed-dotted lines to $M = 2, 4$ mixtures. Insets: the apparent power-law exponents as a function of $f_2$.}
\label{fig:G_omega_mixtures}
\end{figure}

Focussing on the lowest investigated temperature, for which the power-law regime is the longest, we can fit the frequency-dependence of the storage moduli, which display behaviours compatible with power laws in a larger range compared to loss moduli, to obtain an apparent exponent for all the investigated systems. The numerical data for the fitted $G'(\omega)$ are shown in Figure~\ref{fig:G_omega_mixtures}, while the resulting exponents are reported in the insets of the same figure. Quantitatively, the $\lambda = 1$ exponents are always slightly smaller than their $\lambda = 10$ counterparts. By contrast, the qualitative behaviour of the $f_2$ dependence of the exponents is the same for the swapping and non-swapping cases: it is $\approx 0$, which signals solid-like behaviour, for pure $M = 3$ and $M = 4$ systems, and increases monotonically as $f_2$ increases, reaching a value $\approx 0.6$ for pure divalent systems. Such a value is intermediate between $1/2$ and $3/4$, which are the predicted exponents for flexible~\cite{rubinsten2003polymer}, and semiflexible~\cite{PhysRevE.58.R1241} polymers, respectively, and can be explained by noting that divalent patchy particles form living polymers with an exponential size distribution~\cite{sciortino2007self}. Therefore, the system can be seen as a dynamic polydisperse mixture of short (semiflexible) and long (flexible) chains~\cite{10.1063/5.0134271}.

The exponents extracted from the binary mixtures response functions are intermediate between the pure systems, with the divalent-trivalent mixtures displaying an almost linear dependence on $f_2$, compared to the markedly non-linear behaviour of the divalent-tetravalent systems. These results show that the observed power-law behaviour for mixtures of divalent and $M$-valent particles, which can be tuned by the relative concentrations, can be exploited to complement existing routes to rationally designs systems displaying power-law rheology~\cite{conrad2023towards}.

\section{Conclusions}

To summarise, we simulated valence-limited systems at different temperatures, following their evolution from a dense gas to a connected network. We first focussed on tetravalent patchy particles, and then extended the analysis to fluids composed of trivalent and divalent particles, as well as their mixtures. We evaluated the dynamics in terms of diffusion and viscosity, finding that at low temperature all systems display an Arrhenius behaviour with an activation energy that is linked to the average valence and bonding strength. Interestingly, in all cases we found that the Stokes-Einstein relation without a hydrodynamic diameter~\cite{ohtori2018stokes,costigliola2019revisiting,khrapak2021excess} holds, with only a hint of a breakdown at low temperatures. 

Finally, we showed that at low temperature these disordered materials exhibit a viscoelastic response that, intriguingly, generates a power-law behaviour in the frequency-dependence of $G'$ and $G''$ if divalent particles are added to the system, providing a minimal model where power-law rheology can be investigated in associating systems.

A comparison of the numerical results with the available experimental data from literature showed a positive agreement, strengthening the link between the limited-valence model system used here and its available experimental realisations, such as DNA nanostars.

\section*{Acknowledgements}

We acknowledge the CINECA award under the ISCRA initiative, for the availability of high-performance computing resources and support. We thank Francesco Sciortino for useful discussions.

\section*{Data availability}

The code used to run all simulations, together with selected example input files and the data used in the paper can be found in Ref.~\cite{gomez_2024_10838270}.

\section*{Appendix} 

\renewcommand{\thefigure}{A\arabic{figure}}
\setcounter{section}{0}
\setcounter{equation}{0}
\setcounter{figure}{0}

\renewcommand{\thesection}{Appendix\Roman{section}}
\renewcommand{\theequation}{A\arabic{equation}}

\subsection{Mean-squared displacements}
\label{app:msd}

\begin{figure}[ht!]
   \centering
   \includegraphics[width=0.4\textwidth]{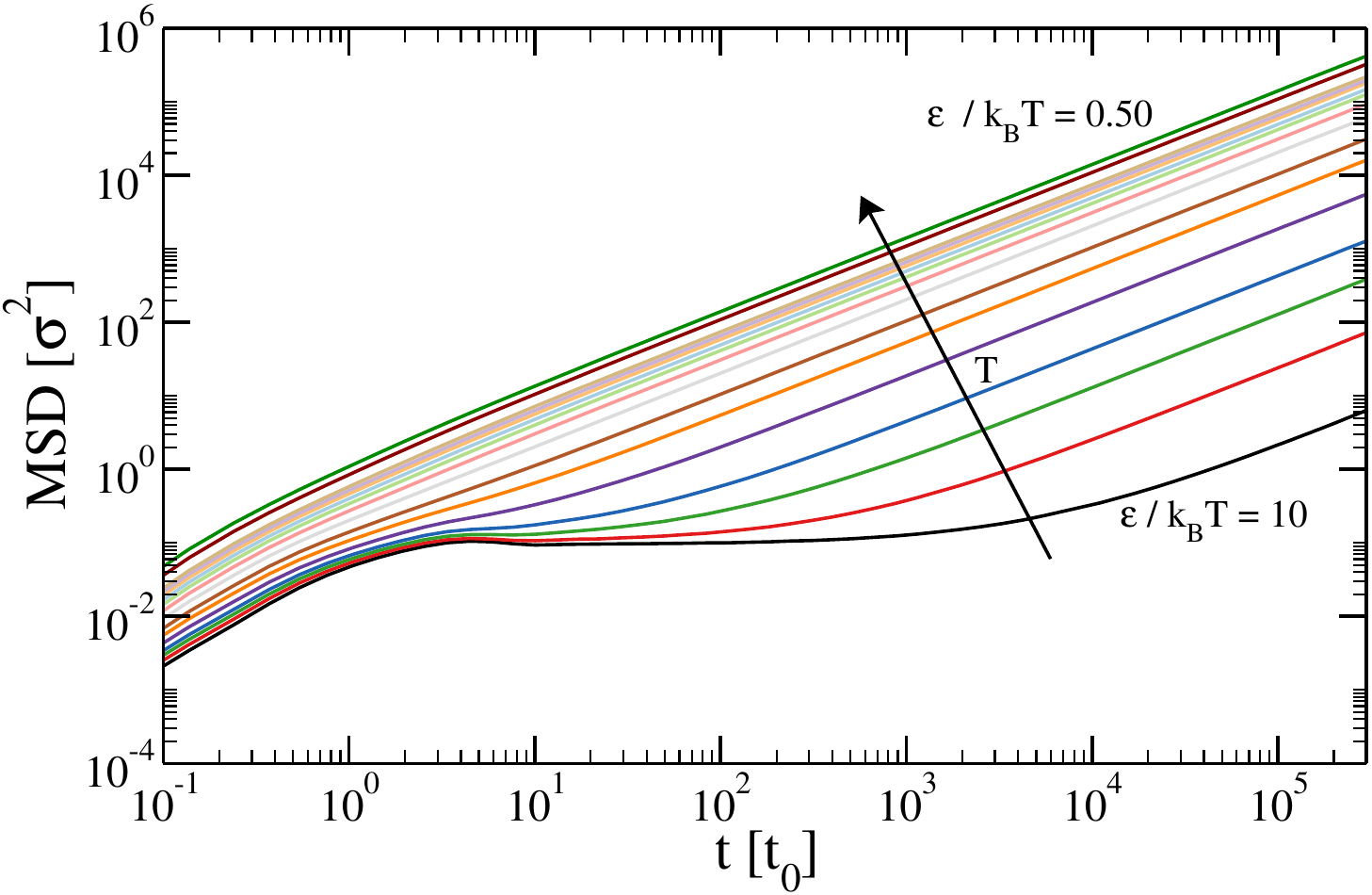} 
   \includegraphics[width=0.4\textwidth]{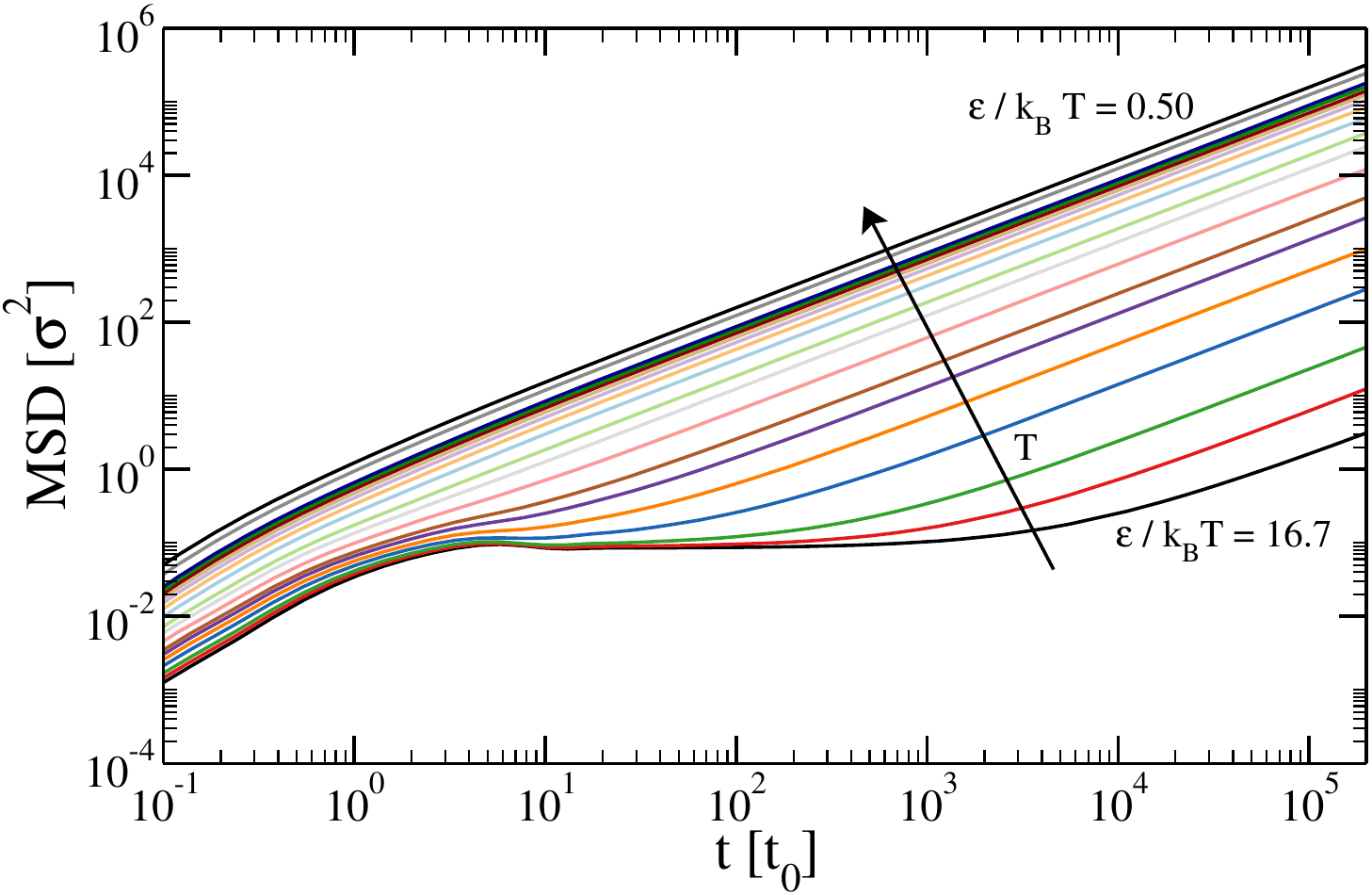} 
   \caption{Mean-squared displacements for the $M = 4$ systems with (a) $\lambda = 10$ and (b) $\lambda = 1$ simulated at $\rho \sigma^3 = 0.60$ for all investigated temperatures.}
\label{fig:M4_msd}
\end{figure}

Figure~\ref{fig:M4_msd} shows the mean-squared displacements of the $M = 4$, $\rho \sigma^3 = 0.60$ swapping and non-swapping systems for different temperatures. At all temperatures the behaviour is ballistic at small times and diffusive at sufficiently large times. As $T$ decreases, a subdiffusive regime appears for intermediate times, developing in a well-resolved plateau that spans two orders of magnitude in time at the lowest temperatures considered. The other systems considered (pure systems with $M = 2$ and $3$, binary mixtures) display the same qualitative trends.

\subsection{Stokes-Einstein relation for $M = 3$ pure and mixture systems}
\label{app:SE_M3}

\begin{figure}[ht!]
   \centering
   \includegraphics[width=0.45\textwidth]{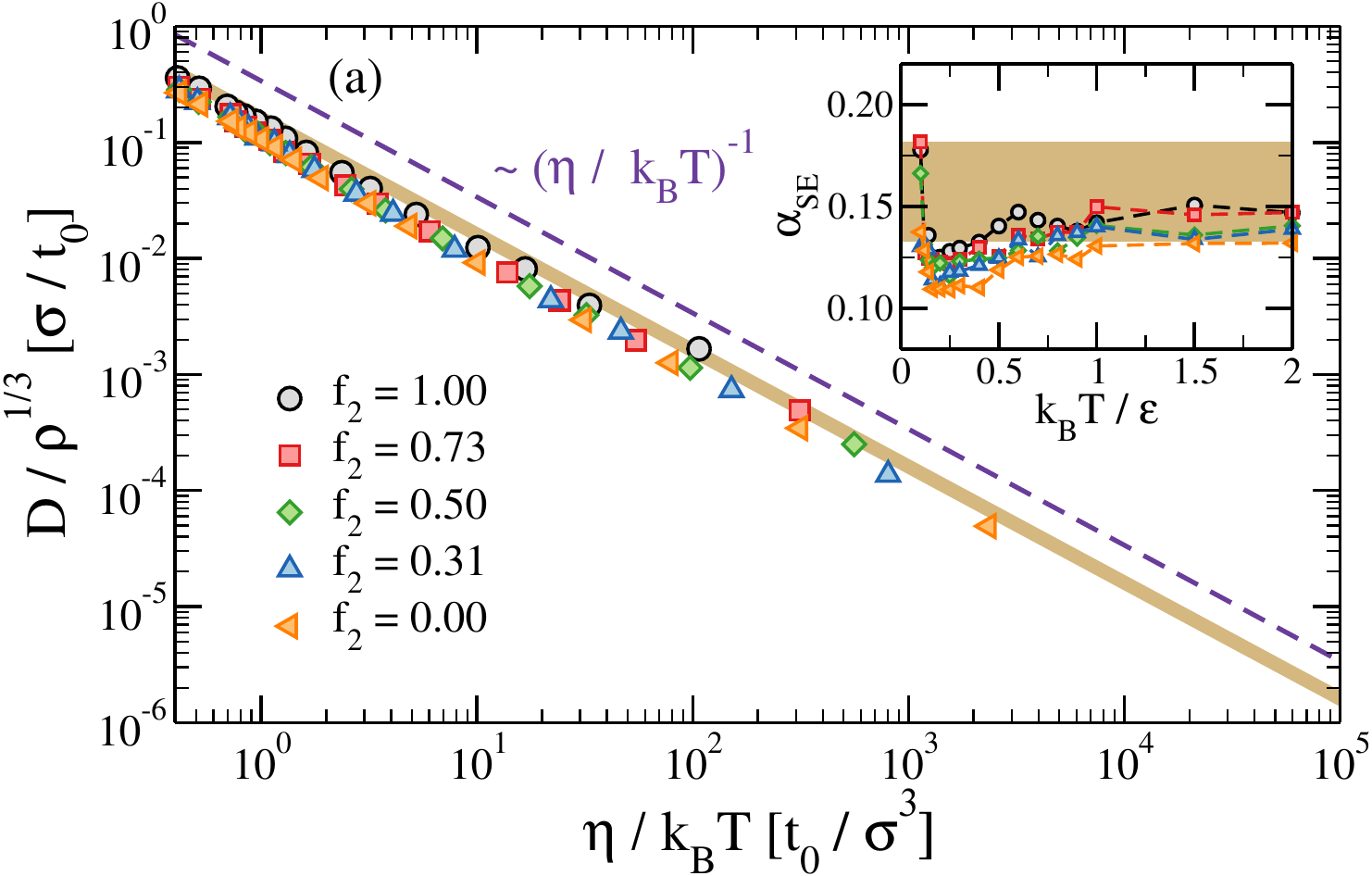} 
   \includegraphics[width=0.45\textwidth]{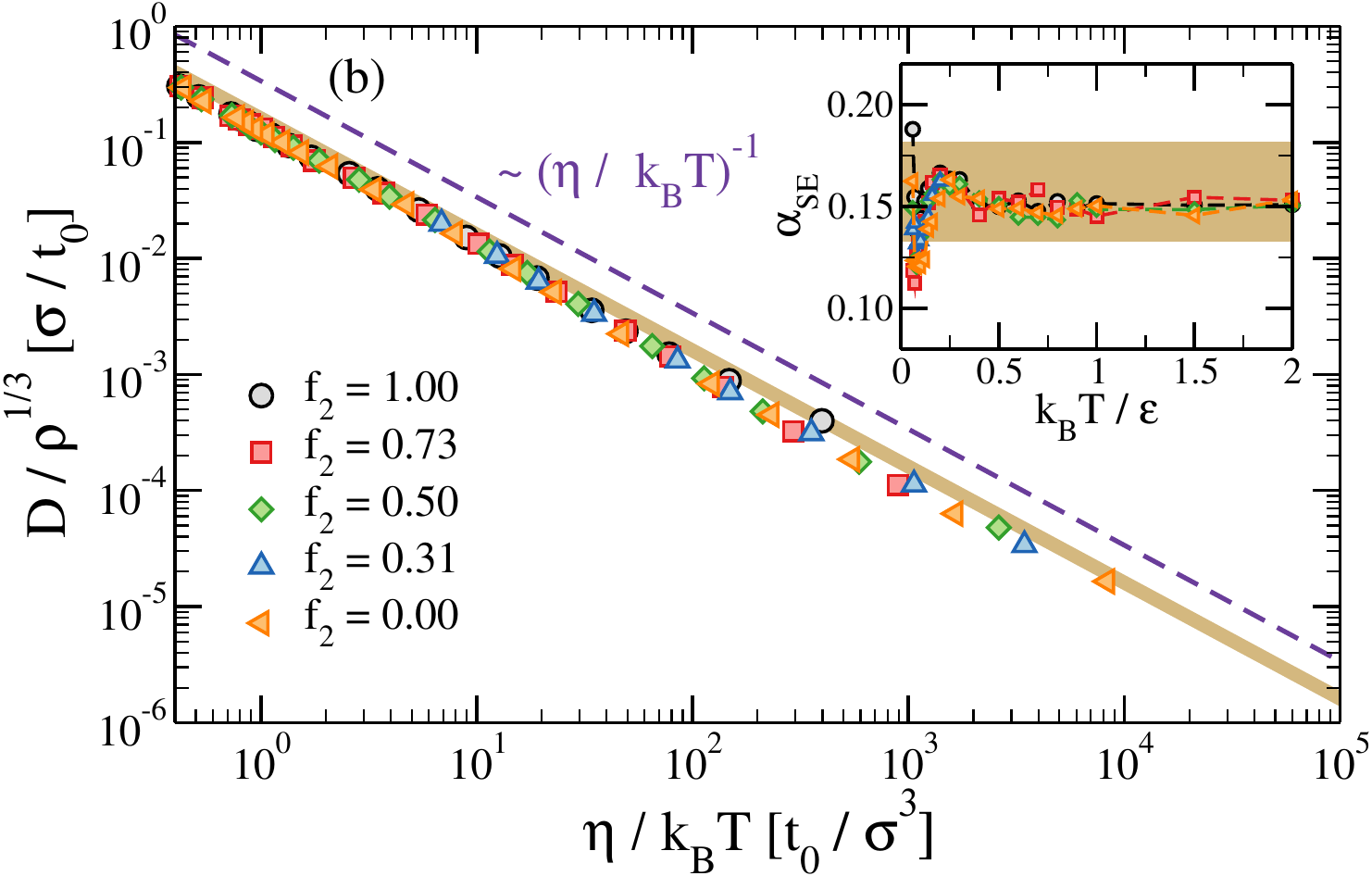} 
   \caption{The diffusion constant scaled by the average interparticle distance, $D / \rho^{1/3}$, as a function of the rescaled viscosity, $\eta / k_B T$, for the (a) non-swapping and (b) swapping systems composed of $M = 3$ particles only ($f_2 = 0$) or mixed with divalent particles ($f_2 > 0$). The violet line is the slope predicted by Eq.~\eqref{eq:SE}, while the shaded brown region bounds the theoretical range as predicted by Zwanzig's theory~\cite{zwanzig1983relation}. The insets show the numerical value of $\alpha_{SE}$, Eq.~\eqref{eq:alpha}, for the same systems.}
\label{fig:M3_SE}
\end{figure}

Figure~\ref{fig:M3_SE} shows a test of the Stokes-Einstein relation without hydrodynamic diameter for systems made of $M = 3$ particles (pure or mixed with divalent particles). From the qualitative point of view, the same qualitative behaviour discussed for the $M = 4$ system and its mixtures is observed (see Fig.~\ref{fig:M4_SE}).

\bibliography{bibliography.bib}

\end{document}